\documentclass[a4paper,fleqn,usenatbib]{mnras}
\usepackage{CJK}
\usepackage{newtxtext,newtxmath}
\usepackage[T1]{fontenc}
\usepackage{ae,aecompl}
\usepackage[]{units}
\usepackage{graphicx}
\usepackage{amssymb}
\usepackage{amsmath}
\usepackage{threeparttable}
\usepackage{hyperref}
\usepackage{color}
\usepackage{hhline}
\usepackage{subfigure}
\usepackage{lineno}

\newcommand{\bpg}{b_{p,g}}
\newcommand{\qeff}{Q_{\rm{eff}}}

\mathchardef\mhyphen="2D

\title[3PCF in squeezed limit]{Using galaxy pairs to investigate the three-point correlation function in the squeezed limit}

\author[S. Yuan, D. J. Eisenstein and L. H. Garrison]{
Sihan Yuan,$^{1}$\thanks{E-mail: sihan.yuan@cfa.harvard.edu}
Daniel J. Eisenstein,$^{1}$
and Lehman H. Garrison$^{1}$
\\
$^{1}$Harvard-Smithsonian Center for Astrophysics, 60 Garden St., Cambridge, MA 02138, USA
}

\date{Accepted XXX. Received YYY; in original form ZZZ}

\pubyear{2017}

\begin{document}
\label{firstpage}
\pagerange{\pageref{firstpage}--\pageref{lastpage}}
\maketitle

\begin{abstract} 

We investigate the three-point correlation function (3PCF) in the squeezed limit by considering galaxy pairs as discrete objects and cross-correlating them with the galaxy field. 
We develop an efficient algorithm using Fast Fourier Transforms to compute such cross-correlations and their associated pair-galaxy bias $b_{p,g}$ and the squeezed 3PCF coefficient $\qeff$.
We implement our method using N-body cosmological simulations and a fiducial Halo Occupation Distribution (HOD) and present the results in both the real space and redshift space. In real space, we observe a peak in $b_{p,g}$ and $\qeff$ at pair separation of $\sim$~2 Mpc, attributed to the fact that galaxy pairs at 2 Mpc separation trace the most massive dark matter halos. We also see strong anisotropy in the $b_{p,g}$ and $\qeff$ signals that track the large-scale filamentary structure. In redshift space, both the 2~Mpc peak and the anisotropy are significantly smeared out along the line-of-sight due to Finger-of-God effect. In both the real space and redshift space, the squeezed 3PCF shows a factor of 2 variation, contradicting the hierarchical ansatz but offering rich information on the galaxy-halo connection. 
Thus, we explore the possibility of using the squeezed 3PCF to constrain the HOD. When we compare two simple HOD models that are closely matched in their projected two-point correlation function (2PCF), we do not yet see a strong variation in the 3PCF that is clearly disentangled from variations in the projected 2PCF. Nevertheless, we propose that more complicated HOD models, e.g. those incorporating assembly bias, can break degeneracies in the 2PCF and show a distinguishable squeezed 3PCF signal. 

\end{abstract}
\begin{keywords}
cosmology: large-scale structure of Universe -- cosmology: dark matter -- galaxies: haloes -- methods: analytical 
\end{keywords}
\section{Introduction}
\label{sec:intro}

Understanding galaxy clustering is one of the primary goals of cosmology, and the advent of a new generation of galaxy surveys has ushered in the new era of high precision galaxy clustering studies.
 A key objective of such studies is to understand the relationship between galaxies and the underlying dark matter halos since such relationship is critical in helping us infer the linear dark matter power spectrum from the observed galaxy power spectrum. The Halo Occupation Distribution (HOD) is the most popular framework for describing how the number of galaxies per halo depends the halo mass \citep{2000Seljak, 2000Peacock, 2001Scoccimarro, 2002Berlind, 2002Cooray, 2004Kravtsov, 2005Zheng}. For the simple HOD we consider in this paper, galaxies populating a halo are divided into central and satellite galaxies, hereafter referred to simply as centrals and satellites \citep{2003Berlind, 2005Zheng}.

There has been a wealth of studies that utilize the two-point correlation functions (2PCF) or higher-order statistics to constrain the HOD \citep[e.g.][]{2005Zehavi, 2007Kulkarni, 2008Blake, 2009Zheng, 2011White, 2013Parejko, 2015More, 2016Guo, 2016Rodriguez}. 
 Another popular way to constrain the HOD is by probing the halo mass of galaxies via weak gravitational lensing, commonly know as galaxy-galaxy lensing, \citep{2006Mandelbaum, 2006Yoo, 2009Li, 2012Leauthaud, 2012Tinker, 2013Mandelbaum, 2015Miyatake, 2015More}. 
Other methods for constraining the HOD include a pair-counting based technique called the Counts-in-Cylinders (CiC) \citep{2009Reid}, and direct counting of the number of galaxies in clusters as a function of halo mass \citep{2009Ho, 2015Hoshino}. 

For this paper, we are interested in constraining the high mass regime of the HOD ($M\sim 10^{14-15} M_{\odot}$), where each dark matter halo hosts multiple galaxies close together. This suggests that if we consider close pairs of galaxies as objects in their own right, they would be biased tracers of high mass halos (refer to Figure~\ref{fig:mhalo_hist}).

The first step is to formulate the proper statistic on close galaxy pairs to calculate their bias. One might choose to use the pair-pair auto-correlation function. However, to preserve the pair separation dependence of our statistic, we need to auto-correlate much smaller subsamples of pairs binned by pair separation. Auto-correlating smaller samples leads to a noisier statistic. 
Thus, instead of auto-correlating close pairs, we investigate the cross-correlation of close pairs with galaxy singlets to calculate the pair-galaxy cross-correlation function. The large-scale pair-galaxy cross-correlation then reveals the linear bias \citep{1996Fry, 1998Tegmark, 1998Scherrer} of the pair field relative to the galaxy field, i.e. the pair-galaxy bias, which we denote as $\bpg$, given by
\begin{equation}
b_{p,g} = \xi_{p,g}/\xi_{g,g},
\end{equation} 
where $\xi_{p,g}$ and $\xi_{g,g}$ are the pair-galaxy cross-correlation and the galaxy-galaxy cross-correlation, respectively. When the pair separation is much less than the distance to the galaxy singlet, the pair-galaxy cross-correlation is equivalent to the three-point correlation function (3PCF) in the squeezed limit. For conciseness, we refer to the 3PCF in the squeezed limit as the squeezed 3PCF in the rest of the paper.

Generally speaking, the 3PCF, $\zeta(\textbf{r}_1, \textbf{r}_2, \textbf{r}_3)$, describes the probability distribution of finding three galaxies forming a triangle with three sides given by vectors $\textbf{r}_1$, $\textbf{r}_2$, and $\textbf{r}_3$ (See \citet{2002Bernardeau} for a comprehensive review). In the hierarchy of correlation functions, the 3PCF is the second lowest order correlation function after the 2PCF, and is the lowest order correlation function to provide shape information \citep{1980Peebles}.
The squeezed limit of the 3PCF refers to the limit where one side of the triangle is much shorter than the other two sides, i.e. $r_1 \ll r_2\approx r_3$. 


The spatial distribution of matter in the Universe can be fully described by the 2PCF if the matter density field is fully Gaussian. However, we expect
non-linear gravitational evolution to produce non-Gaussian signatures in the galaxy distribution that we measure today \citep{2002Bernardeau}. The 3PCF is thus needed to complement the information provided by the 2PCF when describing late-time galaxy distribution.
 A series of recent papers have focused on measuring and analysing the 3PCF in modern cosmological surveys, such as \citet{2009Gaztanaga, 2011Marin, 2015Gil-Marin, 2017bSlepian, 2017aSlepian}. The additional information provided by the 3PCF is often used to break model degeneracies describing cosmology and galaxy bias \citep{1994Gaztanaga, 2004Jing, 2007Kulkarni, 2007Zheng, 2011McBride}. 
 Additionally, the 3PCF and its Fourier space counterpart, the bispectrum, are also commonly used to probe primordial non-gaussianities as a test of single-field slow-roll inflationary models \citep[e.g.][]{2008Dalal, 2009Fergusson, 2010Sefusatti, 2011Baldauf, 2014Tasinato, 2016Hashimoto}. \citet{2016Tellarini} offers a nice review of this topic. 
 

Measuring the 3PCF is computationally intense as the number of triplets scales as $N^3$ and thus becomes prohibitive for large samples of galaxies.
The 3PCF in the squeezed limit is advantageous because its computation can be reduced to two separate pair finding problems of complexity $O(N^2)$. Thus, even though we phrased our statistic as a cross-correlation between the pair field and the galaxy field, the same statistic can be interpreted as a squeezed 3PCF. We show the equivalence of the two approaches in Section~\ref{sec:bpg_to_Qeff}.


In the following sections of the paper, we present a fast and efficient way of calculating the pair-galaxy bias using Fast Fourier Transforms (FFTs) in Section~\ref{sec:bpg}, and how it translates to the squeezed 3PCF in Section~\ref{sec:bpg_to_Qeff}. We discuss implementing our methodology for Luminous Red Galaxies (LRGs) using N-body cosmological simulations in Section~\ref{sec:implement}, and present the results in Section~\ref{sec:results}. In Section~\ref{sec:hod_depend}, we discuss the possibility of constraining HOD with the pair-galaxy bias and the squeezed 3PCF. Finally, we draw a few concluding remarks in Section~\ref{sec:conclusions}.

\section{Methodology}
\label{sec:method}

In this section, we develop a fast and efficient method of computing $\bpg$ using FFT and then formulate the exact relationship between $\bpg$ and the squeezed 3PCF starting from the triplet counting definition of the 3PCF. 

\subsection{Computing pair-galaxy bias $\protect\bpg$}
\label{sec:bpg}

Consider a galaxy overdensity field $\delta_g(\textbf{x})$, a galaxy pair overdensity field $\delta_p(\textbf{x})$, and their Fourier space counterparts $\tilde{\delta_g}(\textbf{k})$ and $\tilde{\delta_p}(\textbf{k})$. Assuming white noise denoted by $\sigma$, for a single wave vector $\textbf{k}$, we can write out a $\chi^2$ objective function
\begin{equation}
\chi^2 = \frac{|\tilde{\delta_p}(\textbf{k}) - b_{p,g}\tilde{\delta_g}(\textbf{k})|^2}{\sigma^2}.
\label{equ:chi2}
\end{equation}
Then by minimizing $\chi^2$ against $b_{p,g}$, 
we find a best fit for $b_{p,g}$ and its uncertainties at wave vector $\textbf{k}$
\begin{equation}
b_{p,g}(\textbf{k}) = \frac{\textrm{Re}(\tilde{\delta_g}^*(\textbf{k})\tilde{\delta_p}(\textbf{k}))}{\tilde{\delta_g}^*(\textbf{k})\tilde{\delta_g}(\textbf{k})}, \ \ \ \ 
\sigma_b^{-2}(\textbf{k}) = \frac{\tilde{\delta_g}^*(\textbf{k})\tilde{\delta_g}(\textbf{k})}{\sigma^2}.
\label{equ:bmin}
\end{equation}
Then we average over a range of $k$ modes, weighted by inverse variance multiplied with some weighting function $\tilde{W}(k)$. We get
\begin{align}
b_{p,g} & = \frac{\sum_{\textbf{k}} b_{p,g}(\textbf{k}) \sigma_b^{-2}(\textbf{k}) \tilde{W}(\textbf{k})}{\sum_{\textbf{k}} \sigma_b^{-2}(\textbf{k}) \tilde{W}(\textbf{k})} \nonumber \\ 
& = \frac{\sum_{\textbf{k}} \textrm{Re}(\tilde{\delta_g}^*(\textbf{k})\tilde{\delta_p}(\textbf{k})) \tilde{W}(\textbf{k})}{\sum_{\textbf{k}} \tilde{\delta_g}^*(\textbf{k})\tilde{\delta_g}(\textbf{k}) \tilde{W}(\textbf{k})} \nonumber \\ 
& = \frac{\int d^3 \textbf{k} \  \tilde{W}(\textbf{k}) \textrm{Re}(\tilde{\delta_g}^*(\textbf{k})\tilde{\delta_p}(\textbf{k}))}{\int d^3 \textbf{k} \  \tilde{W}(\textbf{k}) P_g(\textbf{k})}, \label{equ:bpg_initial}
\end{align}
where $P_g(\textbf{k}) = \tilde{\delta_g}^*(\textbf{k})\tilde{\delta_g}(\textbf{k})$ is the galaxy power spectrum. A detailed discussion on the choice of weighting function $\tilde{W}(k)$ is provided in Section~\ref{sec:weighting}. 

Equation~\ref{equ:bpg_initial} shows that $\bpg$ is linear in $\delta_p$, which means that we can compute the integrals for a single pair at a time and then sum over all pairs to obtain $\bpg$.  For the $i$-th galaxy pair, we can define its position $\textbf{x}_{p,i}$ as the middle of the pair. Then the fractional overdensity function of pairs is given by
\begin{equation}
\delta_{p, i}(\textbf{x}) = \frac{\delta_D(\textbf{x}-\textbf{x}_{p,i})}{\bar{n}_p} - 1,
\label{equ:delta_pi_x}
\end{equation}
where $\delta_D(\textbf{x})$ is the Dirac delta function. $\bar{n}_p$ denotes the mean number density of galaxy pairs. The Fourier Transform is given by
\begin{equation}
\tilde{\delta}_{p, i}(\textbf{k}) = \frac{e^{-i\textbf{k}\cdot \textbf{x}_{p,i}}}{\bar{n}_p} - \frac{\delta_D(\textbf{k})}{(2\pi)^3}
\label{equ:delta_pi_k}
\end{equation}
Then we calculate the value $\textrm{Re}(\tilde{\delta_g}^*(\textbf{k})\tilde{\delta_p}(\textbf{k}))$ in equation~\ref{equ:bpg_initial} as
\begin{align}
&\textrm{Re}(\tilde{\delta_g}^*(\textbf{k})\tilde{\delta_p}(\textbf{k}))  = \textrm{Re}(\tilde{\delta_g}^*(\textbf{k})\sum_i\tilde{\delta}_{p,i}(\textbf{k})) \nonumber \\
& = \frac{1}{2} \sum_i (\tilde{\delta_g}^*(\textbf{k}) \tilde{\delta}_{p,i}(\textbf{k}) + \tilde{\delta_g}(\textbf{k}) \tilde{\delta}_{p,i}^*(\textbf{k})) \nonumber \\
& = \frac{1}{2\bar{n}_p} \sum_i (\tilde{\delta_g}^*(\textbf{k}) e^{-i\textbf{k}\cdot \textbf{x}_{p,i}} + \tilde{\delta_g}(\textbf{k}) e^{i\textbf{k}\cdot \textbf{x}_{p,i}}) \nonumber \\
& \ \ \ - \frac{N_p}{2} \frac{\delta_D(\textbf{k})}{(2\pi)^3}(\tilde{\delta_g}(\textbf{k}) + \tilde{\delta_g}^*(\textbf{k})), \label{equ:Re_del_del}
\end{align}
where $N_p$ is the total number of galaxy pairs. When we plug the second term of equation~\ref{equ:Re_del_del} into equation~\ref{equ:bpg_initial}, we get a result proportional to $(\tilde{\delta_g}(\textbf{0}) + \tilde{\delta_g}^*(\textbf{0}))$, which gives
\begin{equation}
\tilde{\delta_g}(\textbf{0}) = \tilde{\delta_g}^*(\textbf{0}) = \int d^3 \textbf{x}\  \delta_g(\textbf{x}) = 0. 
\end{equation}
Thus, the second term of equation~\ref{equ:Re_del_del} does not contribute to the numerator in equation~\ref{equ:bpg_initial}. 

By combining equation~\ref{equ:Re_del_del} and equation~\ref{equ:bpg_initial}, we get
\begin{align}
& b_{p,g} \left[ \int d^3 \textbf{k} \  \tilde{W}(\textbf{k}) P_g(\textbf{k})\right] \nonumber \\
& = \frac{1}{2\bar{n}_p} \sum_i \left[\int d^3\textbf{k}\  \tilde{W}(\textbf{k}) (\tilde{\delta_g}^*(\textbf{k}) e^{-i\textbf{k}\cdot \textbf{x}_{p,i}} + \tilde{\delta_g}(\textbf{k}) e^{i\textbf{k}\cdot \textbf{x}_{p,i}})\right] \nonumber \\
& = \frac{1}{2\bar{n}_p} \sum_i \left[\int d^3\textbf{k}\  \tilde{W}(-\textbf{k}) \tilde{\delta_g}(-\textbf{k}) e^{-i\textbf{k}\cdot \textbf{x}_{p,i}}\right.  \nonumber \\
& \ \ \ \ \ \ \ \ \  + \left. \int d^3\textbf{k}\  \tilde{W}(\textbf{k}) \tilde{\delta_g}(\textbf{k}) e^{i\textbf{k}\cdot \textbf{x}_{p,i}}\right] \nonumber \\
& = \frac{1}{\bar{n}_p} \sum_i \left[\int d^3\textbf{k}\  \tilde{W}(\textbf{k}) \tilde{\delta_g}(\textbf{k}) e^{i\textbf{k}\cdot \textbf{x}_{p,i}}\right] \nonumber \\
& = \frac{1}{\bar{n}_p}\sum_i (W \ast \delta_g)(\textbf{x}_{p,i}). \label{equ:bpg_intermediate}
\end{align}
Note that during this derivation, we invoke that $W(\textbf{x})$ is even and real and that $\delta_g(\textbf{x})$ is real. Similarly, we can evaluate the denominator of equation~\ref{equ:bpg_initial} 
\begin{equation}
\int d^3 \textbf{k} \  \tilde{W}(\textbf{k}) P_g(\textbf{k}) = \frac{1}{\bar{n}_g} \sum_j (W \ast \delta_g)(\textbf{x}_{g,j}),
\label{equ: bpg_denominator}
\end{equation}
where we sum over all the galaxy positions $\textbf{x}_{g,j}$, and $\bar{n}_g$ is the mean galaxy number density. 

Thus, we get the final expression for pair-galaxy bias
\begin{equation}
b_{p,g} = \frac{\frac{1}{\bar{n}_p}\sum_i (W \ast \delta_g)(\textbf{x}_{p,i})}{\frac{1}{\bar{n}_g} \sum_j (W \ast \delta_g)(\textbf{x}_{g,j})}.
\label{equ:bpg_final}
\end{equation}
This formula gives a fast and efficient way of calculating $\bpg$ in linear time as a function of the number of galaxy pairs. The numerator allows us to compute only one convolution for the full pair sample and simply evaluate it at the positions of any subsample of pairs whereas the denominator also only requires one convolution. This is much faster than explicitly cross-correlating each subsample of pairs with the overall galaxy field. 

\subsection{Relating $\protect\bpg$ to the the squeezed limit of the 3PCF}
\label{sec:bpg_to_Qeff}

To relate $\bpg$ to the squeezed 3PCF, we first evaluate the denominator of equation~\ref{equ:bpg_final} 
\begin{align}
& \frac{1}{\bar{n}_g} \sum_j (W \ast \delta_g)(\textbf{x}_{g,j}) \nonumber \\
&  =  \frac{1}{\bar{n}_g} \sum_j \int d^3\textbf{x}'\  \delta_g(\textbf{x}' + \textbf{x}_{g,j})W(\textbf{x}') \nonumber \\
& = \frac{1}{\bar{n}_g} \sum_j \int d^3\textbf{x}'\  \left(\frac{n_g(\textbf{x}' + \textbf{x}_{g,j}|\textbf{x}_{g,j})}{\bar{n}_g} - 1\right)W(\textbf{x}'),  
\label{equ:eval_denom}
\end{align}
where $n_g(\textbf{x}' + \textbf{x}_{g,j}|\textbf{x}_{g,j})$ is the number of galaxies at $\textbf{x}' + \textbf{x}_{g,j}$ given an arbitrary galaxy at $\textbf{x}_{g,j}$. By definition of 2PCF $\xi(\textbf{x}')$, we have 
\begin{equation}
n_g(\textbf{x}' + \textbf{x}_{g,j}|\textbf{x}_{g,j}) = \bar{n}_g(1+\xi(\textbf{x}'))
\label{equ:2PCF}
\end{equation}
Thus, combining equation~\ref{equ:eval_denom} and equation~\ref{equ:2PCF}, we have 
\begin{equation}
\frac{1}{\bar{n}_g} \sum_j (W \ast \delta_g)(\textbf{x}_{g,j})  = \int d^3\textbf{x}'\  \xi(\textbf{x}')W(\textbf{x}').
 \label{equ:bpg_numerator}
\end{equation}

Similarly to equation~\ref{equ:eval_denom}, we can evaluate the numerator of equation~\ref{equ:bpg_final} as
\begin{align}
& \frac{1}{\bar{n}_p}\sum_i (W \ast \delta_g)(\textbf{x}_{p,i}) \nonumber \\
&= \frac{1}{\bar{n}_p} \sum_i \int d^3\textbf{x}'\  \left(\frac{n_g(\textbf{x}' + \textbf{x}_{p,i}|\textbf{x}_{p,i})}{\bar{n}_g} - 1\right)W(\textbf{x}'),
\label{equ:denom_intermediate}
\end{align}
where $n_g(\textbf{x}' + \textbf{x}_{p,i}|\textbf{x}_{p,i})$ is the number of galaxies at $\textbf{x}' + \textbf{x}_{p,i}$ given an arbitrary galaxy pair at position $\textbf{x}_{p,i}$.  This number relates to the 3PCF by the following
\begin{align}
& n_g(\textbf{x}' + \textbf{x}_{p,i}|\textbf{x}_{p,i}) = \bar{n}_g(1+\xi(\textbf{d}_p) + \xi(\textbf{x}'+\frac{\textbf{d}_p}{2}) \nonumber \\
& + \xi(\textbf{x}'-\frac{\textbf{d}_p}{2}) + \zeta(\textbf{d}_p, \textbf{x}'+\frac{\textbf{d}_p}{2}, \textbf{x}'-\frac{\textbf{d}_p}{2}))/(1+\xi(\textbf{d}_p)),  
\label{equ:3pf}
\end{align}
where $\textbf{d}_p$ is the separation vector of the galaxy pair. 
Now we combine equation~\ref{equ:denom_intermediate} and equation~\ref{equ:3pf} in the squeezed limit. Then, we have 
\begin{align}
&\frac{1}{\bar{n}_p}\sum_i (W \ast \delta_g)(\textbf{x}_{p,i}) \nonumber \\
& = \frac{1}{\bar{n}_p} \sum_i \int d^3\textbf{x}'\  \left(\frac{\xi(\textbf{x}'_1) +\xi(\textbf{x}'_2) + \zeta(\textbf{d}_p, \textbf{x}'_1, \textbf{x}'_2)}{1+\xi(\textbf{d}_p)}\right)W(\textbf{x}')
\end{align}
where we have defined $\textbf{x}'_1 = \textbf{x}'+\textbf{d}_p/2$ and $\textbf{x}'_2 = \textbf{x}'-\textbf{d}_p/2$.

For our purpose, we are only summing over pairs within a specific $\textbf{d}_p$ bin. Thus, to good approximation, we can treat $\textbf{d}_p$ as approximately constant in the summation, which removes all dependence on pair position inside the summation. Then we have 
\begin{align}
& \frac{1}{\bar{n}_p}\sum_i (W \ast \delta_g)(\textbf{x}_{p,i})  \nonumber \\
& \ \ \  =  \frac{\int d^3\textbf{x}'W(\textbf{x}')(\xi(\textbf{x}'_1)+\xi(\textbf{x}'_2))}{1+\xi(\textbf{d}_p)}+ \frac{\int d^3\textbf{x}' W(\textbf{x}')\zeta(\textbf{d}_p, \textbf{x}'_1, \textbf{x}'_2)}{1+\xi(\textbf{d}_p)}.
\label{equ:bpg_numerator_final}
\end{align}
Then, using equation~\ref{equ: bpg_denominator}, equation~\ref{equ:bpg_final}, and equation~\ref{equ:bpg_numerator_final}, we can express the pair-galaxy bias in terms of the 2PCF and the 3PCF
\begin{align}
b_{p,g} = \frac{A + B}{1+\xi(\textbf{d}_p)},
\label{equ:bpg_3pf}
\end{align}
where
\begin{align}
A & = \frac{\int d^3\textbf{x}'W(\textbf{x}')(\xi(\textbf{x}'_1)+\xi(\textbf{x}'_2))}{\int d^3\textbf{x}'\ W(\textbf{x}') \xi(\textbf{x}')}, \nonumber \\
B & = \frac{\int d^3\textbf{x}'W(\textbf{x}')\zeta(\textbf{d}_p, \textbf{x}'_1, \textbf{x}'_2)}{\int d^3\textbf{x}'\  W(\textbf{x}')\xi(\textbf{x}')}. \nonumber
\end{align}
So now we have expressed the pair-galaxy bias as given by equation~\ref{equ:bpg_final} in terms of general case 2PCFs and 3PCFs. 

To proceed, we introduce the reduced 3PCF $Q_{\rm{eff}}$, defined by
\begin{equation}
Q_{\rm{eff}}(\textbf{x}_1,\textbf{x}_2, \textbf{x}_3) = \frac{\zeta(\textbf{x}_1,\textbf{x}_2, \textbf{x}_3)}{\xi(\textbf{x}_1)\xi(\textbf{x}_2)+\xi(\textbf{x}_1)\xi(\textbf{x}_3) + \xi(\textbf{x}_2)\xi(\textbf{x}_3)},
\label{equ:def_Q}
\end{equation} 
where $\textbf{x}_1,\textbf{x}_2, \textbf{x}_3$ are the three separation vectors for any 3PCF. In the original hierarchical model \citep{1975Peebles, 1976Soneira, 1977Groth, 1980Peebles}, the $Q_{\rm{eff}}$ factor is a constant, which has been disproven in later surveys and simulations \citep[e.g.][]{2007Kulkarni, 2008Marin, 2011McBride}. As we will see, our results in the squeezed limit also show a strongly variable $\qeff$.

In the strictly squeezed limit, $|\textbf{x}'| \gg |\textbf{d}_p|$ and $\xi(\textbf{x}'_1)\approx \xi(\textbf{x}'_2) \approx \xi(\textbf{x}')$. A typical 2PCF assumes a power-law form $\xi(\textbf{x}') \propto x'^{\alpha}$, where the exponent $\alpha \approx -2$. This means that, in the strictly squeezed limit, $\xi(\textbf{x}') \ll \xi(\textbf{d}_p)$, and thus
\begin{align}
\zeta(\textbf{d}_p, \textbf{x}'_1, \textbf{x}'_2) & \approx Q_{\rm{eff}}(2\xi(\textbf{x}')\xi( \textbf{d}_p)+\xi(\textbf{x}')^2) \nonumber \\
& \approx 2Q_{\rm{eff}}\xi(\textbf{x}')\xi(\textbf{d}_p).
\label{equ:3pf_hierar}
\end{align}
Then combining equation~\ref{equ:bpg_3pf} and equation~\ref{equ:3pf_hierar}, we express pair-galaxy bias as
\begin{equation}
b_{p,g} = \frac{2(1+Q_{\rm{eff}}\xi(\textbf{d}_p))}{1+\xi(\textbf{d}_p)}.
\label{equ:bpg_squeezed}
\end{equation}
Then we can express the reduced 3PCF $Q_{\rm eff}$ as a function of pair-galaxy bias $\bpg$
\begin{equation}
Q_{\rm eff} = \frac{(1+\xi(\textbf{d}_p))b_{p,g}-2}{2\xi(\textbf{d}_p)}.
\label{equ:Qeff}
\end{equation}
$\qeff$ is not a function of $\textbf{x}'$ as a manifestation of the local bias limit, i.e. the pair separation scale is much smaller than the separation to the third galaxy so that $\qeff$'s dependence on $\textbf{x}'$ converges to scale with $\xi(\textbf{x}')$. Also note that when pair separation is extremely small, i.e. $|\textbf{d}_p| \sim 1$~Mpc, we have $\xi(\textbf{d}_p) \gg 1$, and $Q_{\rm eff}$ would tend to $b_{p,g}/2$. 

However, as we will see in Section~\ref{sec:implement}, the pairs we are considering have a maximum separation of up to 10 or 30 Mpc, where the $|\textbf{x}'| \gg |\textbf{d}_p|$ limit does not always hold. Specifically, we can no longer assume $\xi(\textbf{x}'_1)\approx \xi(\textbf{x}'_2) \approx \xi(\textbf{x}')$ and also the $\xi(\textbf{x}'_1)\xi(\textbf{x}'_2)$ term contributes non-trivially to equation~\ref{equ:3pf_hierar}. We use the general form of equation~\ref{equ:def_Q}
\begin{align}
& \zeta(\textbf{d}_p, \textbf{x}'_1, \textbf{x}'_2) = \nonumber \\
& \ \ \ Q_{\rm{eff}}\left[\xi(\textbf{x}'_1)\xi( \textbf{d}_p)+\xi(\textbf{x}'_2)\xi( \textbf{d}_p)+\xi(\textbf{x}'_1)\xi(\textbf{x}'_2)\right].
\end{align}
Assuming $\xi(r) \propto r^{-2}$ and treating $d_p/x'$ as a small number, we Taylor expand the first two terms, $\xi(\textbf{x}'_1) \xi(\textbf{d}_p)$ and $\xi(\textbf{x}'_2) \xi(\textbf{d}_p)$, to the fourth order $O\left[\left(d_p/x'\right)^4\right]$. We also preserve the zero-th order approximation of the Taylor expansion of the last term $\xi(\textbf{x}'_1)\xi(\textbf{x}'_2) \approx \xi(\textbf{x}')^2 \propto (d_p/x')^4$. 
The Taylor expanded form of the first two terms still carries a dependence on the angle between $\textbf{x}'$ and $\textbf{d}_p$. 
As a rough estimate, we average over the angle between $\textbf{x}'$ and $\textbf{d}_p$ assuming isotropy, we get
\begin{equation}
\xi(\textbf{x}'_1)+\xi(\textbf{x}'_2) \approx 2\xi(\textbf{x}') + \frac{\xi(\textbf{x}')^2}{2\xi(\textbf{d}_p)},
\end{equation}
and
\begin{equation}
\zeta(\textbf{d}_p, \textbf{x}'_1, \textbf{x}'_2) \approx Q_{\rm{eff}}\left[2\xi(\textbf{x}')\xi(\textbf{d}_p)+\frac{3}{2}\xi(\textbf{x}')^2\right].
\end{equation}
Note that since we are assuming isotropy and a simplified form for $\xi(r)$, the coefficient in front of $\xi(\textbf{x}')^2$ is only an estimate of the correction. However, a correction to the correction term comes in as a higher order correction to $\zeta(\textbf{d}_p, \textbf{x}'_1, \textbf{x}'_2)$, so we ignore these high order effects for this derivation. 
Then, in the not strictly squeezed limit, the pair-galaxy bias is given by
\begin{equation}
b_{p,g} = \frac{2(1+Q_{\rm{eff}}\xi(\textbf{d}_p))}{1+\xi(\textbf{d}_p)} + \delta b.
\label{equ:bpg_notsq}
\end{equation}
where
\begin{equation}
\delta b =  \frac{\frac{1}{2\xi(\textbf{d}_p)} + \frac{3}{2}\qeff}{1+\xi(\textbf{d}_p)} \frac{\int d^3\textbf{x}' W(\textbf{x}')\xi(\textbf{x}')^2}{\int d^3\textbf{x}'W(\textbf{x}')\xi(\textbf{x}')}
\end{equation}

Given a weighting function and a 2PCF, the fraction between the two integrals yields a constant. For ease of notation, we express this term as the 2PCF evaluated at some characteristic separation $d_w$, which is dependent on the weighting function and the shape of the 2PCF. Specifically, we define
\begin{equation}
\xi(d_w) \equiv \frac{\int d^3\textbf{x}' W(\textbf{x}')\xi(\textbf{x}')^2}{\int d^3\textbf{x}'W(\textbf{x}')\xi(\textbf{x}')}.
\label{equ:xi_dw}
\end{equation}
To calculate the characteristic separation, we employ the power-law form of the galaxy-galaxy 2PCF, $\xi(r) \propto r ^{-p}$. Then equation~\ref{equ:bpg_notsq} gives the following formula to calculate $d_w$
\begin{equation}
d_w = \left[\frac{\int d^3\textbf{x}' W(\textbf{x}') |\textbf{x}'|^{-p} }{\int d^3\textbf{x}' W(\textbf{x}') |\textbf{x}'|^{-2p}}\right]^{1/p}.
\label{equ:dw}
\end{equation}
Having introduced $\xi(d_w)$, the pair-galaxy bias can then be conveniently expressed with
\begin{equation}
b_{p,g} = \frac{2+\frac{\xi(d_w)}{2\xi(\textbf{d}_p)}+Q_{\rm{eff}}(2\xi(\textbf{d}_p) + \frac{3}{2}\xi(d_w))}{1+\xi(\textbf{d}_p)},
\label{equ:bpg_notsq_dw}
\end{equation}
which translates to a modified estimator of the reduced 3PCF
\begin{equation}
Q_{\rm eff} = \frac{(1+\xi(\textbf{d}_p))b_{p,g}-2-\frac{\xi(d_w)}{2\xi(\textbf{d}_p)}}{2\xi(\textbf{d}_p) + \frac{3}{2}\xi(d_w)}.
\label{equ:Qeff_notsq}
\end{equation}
Thus the reduced 3PCF estimator only differs from the strictly squeezed limit by an extra factor of $\frac{\xi(d_w)}{2\xi(\textbf{d}_p)}$ on the numerator and a factor of $\frac{3}{2}\xi(d_w)$ on the denominator. For the rest of this paper, since we mostly do not operate in the strictly squeezed limit, we use equation~\ref{equ:Qeff_notsq} to calculate $\qeff$.



\section{Implementation}
\label{sec:implement}
In this section, we present the implementation of our $\bpg$ and $\qeff$ statistic for Luminous Red Galaxies (LRGs) using a series of N-body simulation boxes. LRGs are highly luminous early-type massive galaxies consisting of mainly old stars and quenched of star formation. These luminous galaxies are popular targets for large-scale structure studies and surveys. We study the statistics of LRGs specifically in this paper in preparation for applying our methods to Baryon Oscillation Spectroscopic Survey (BOSS; \citealt{2011Eisenstein, 2013Dawson}) data in follow up papers. 

We also present some basic characteristics of the simulation results in this section, namely the 2PCF, pair separation distribution, and halo mass distribution of the LRGs and their pairs. It should be pointed out that we are only interested in showcasing our methodology in this paper, and we do not attempt any detailed matching with observed LRG statistics. 

\subsection{N-body cosmology simulation}
\label{sec:nbody}

For the purpose of this analysis, we utilize the cosmological simulations (Garrison et al., in preparation) generated by the fast and high precision \textsc{Abacus} N-body code (\citealt{2016Garrison}, Ferrer et al., in preparation; Metchnik $\&$ Pinto, in preparation). Specifically we use a series of 16 cyclic boxes with Planck 2016 cosmology \citep{2016Planck} at redshift $z = 0.5$, where each box is of size 1100 $h^{-1}$~Mpc, and contains 1440$^3$ dark matter particles of mass $4\times 10^{10}$ $h^{-1}\ M_{\odot}$.  Dark matter halos are found and characterized using the friends-of-friends (FoF; \citealt{1985Davis}) halo finder. Different halo finders, specifically ROCKSTAR \citep{2013Behroozi}, behave similarly for halos of at least a few hundred particles \citep{2016Garrison}. For the rest of this paper, we use halo catalogues generated using the FoF halo finder with linking length $0.15b$, where $b$ is the average interparticle spacing. We have also repeated the key analysis with mock catalogues based on halos from ROCKSTAR and recovered consistent results.

Although FoF finder is fast, one main concern with FoF finder is the effect of FoF overbridging \citep{2008Lukic, 2011Knebe}, which misidentifies a small fraction of physically separate halos as a single halo through long chain of links. This effect occurs at all scales and eventually leads to a small over-estimate of the number of galaxy pairs inhabiting the same halo. 
For our purpose, the correction is small and would not impact the qualitative results of this paper. 


\subsection{Redshift space distortion}
\label{sec:rsd}

An important effect to consider is Redshift Space Distortion (RSD, \citealt{1998Hamilton}). Essentially, the peculiar velocity of the galaxies in clusters along the line-of-sight (LOS) leads to a deviation in distance measurements when using Hubble's law. This is commonly known as the ``finger-of-god" (FoG) effect. There is also the more subtle effect known as the Kaiser effect \citep{1987Kaiser}, where the coherent infall velocity of galaxies towards central masses introduces a flattening in the observed structure on the large scale. It is important to include RSD in our calculations to give results that are comparable to observables. 
To include RSD, we shift the $x_{\parallel}$ coordinate of each galaxy by
\begin{equation}
x'_{\parallel} = x_{\parallel} + \frac{v_{\parallel}}{H(z)}, 
\label{equ:RSD}
\end{equation}
where $v_{\parallel}$ is the LOS velocity of the galaxy. For simplicity, we set the LOS velocity of the galaxy equal to the LOS velocity of the host dark matter particle, ignoring the $\sim 10\%$ relative velocity bias due to the shear motion between the galaxies and their underlying halos \citep{2015aGuo}. 
Then we can derive the pair-galaxy bias and the squeezed 3PCF in redshift space using the distorted positions of the galaxies. Several previous papers have also studied the behaviors of redshift space 3PCF in the context of galaxy surveys and simulations \citep[e.g.][]{2005Gaztanaga, 2008Marin, 2015bGuo}.

For the rest of this paper, we calculate the squeezed 3PCF in both real space and redshift space. To avoid confusion, we explicit state in the figure caption whether the figures are made in real space or redshift space. 

\subsection{Mock LRG and LRG pair catalogue}
\label{sec:catalogue}

The first step is to generate mock LRG catalogue from the halo catalogue using a HOD prescription. The HOD we use is a slightly modified version of the the simple HOD fitted to the LRG 2PCF in the Sloan Digital Sky Survey (SDSS; \citealt{2000York, 2002Zehavi, 2003Abazajian, 2004Zehavi, 2005bEisenstein, 2005bZehavi}) and summarized in \citet{2009Zheng, 2015Kwan}. The form of our fiducial HOD (equation~\ref{equ:hod}; \citealt{2004Kravtsov, 2005Zheng, 2007bZheng}) is simple in the sense that it only has five parameters, only depends on the halo mass, and ignores effects of halo assembly and halo environment. A brief discussion on more sophisticated halo occupation models is deferred to Section~\ref{sec:3pf_hod}, but for the purpose of providing an simple application of our methodology, this simple HOD that solely depends on halo mass is sufficient \citep[e.g.][]{2004Kauffmann, 2004Mo, 2006Blanton, 2008Tinker}. 

\begin{figure}
\centering
 \hspace*{-0.4cm}
\includegraphics[width=3.6in]{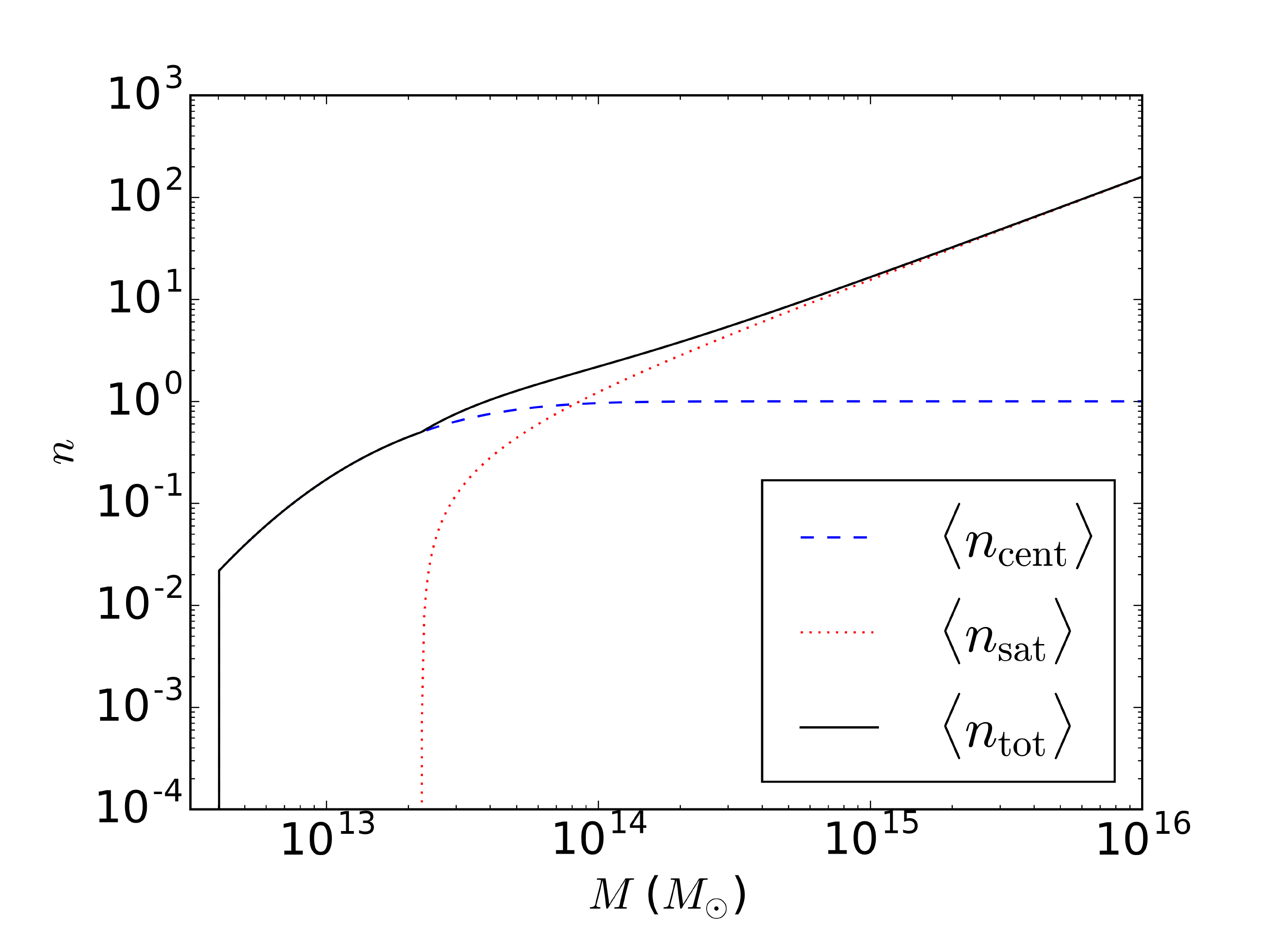}
\vspace{-0.6cm}
\caption{The HOD plotted as number of LRGs per halo for a given halo mass $M$ \citep{2009Zheng, 2015Kwan}. The blue dashed line is the number of centrals whereas the red solid line is the number of satellites. The total number of LRGs per halo is given by the black solid line. }
\label{fig:hod}
\end{figure}

Specifically, this HOD prescription gives the number of centrals and satellites as
\begin{align}
& \langle n_{\mathrm{cent}} \rangle = \frac{1}{2}\mathrm{erfc} \left[\frac{\ln(M_{\mathrm{cut}}/M)}{\sqrt{2}\sigma}\right], \nonumber \\
& \langle n_{\rm{sat}} \rangle = \left(\frac{M-\kappa M_{\rm{cut}}}{M_1}\right)^{\alpha},
\label{equ:hod}
\end{align}
where the parameters are chosen as, $M_{\rm{cut}} \approx 10^{13.35} M_{\odot}$, $M_1 \approx 10^{13.8} M_{\odot}$, $\sigma = 0.85$, $\alpha = 1$, and $\kappa = 1$  \citep{2009Zheng, 2015Kwan}. The halo mass threshold for satellites is given by $M_{\rm{min}} = \kappa M_{\rm cut}$. This HOD is plotted in Figure~\ref{fig:hod}. From here on, we refer to this HOD as the Z09 design. Also note that we truncate the HOD at $M=4\times 10^{12} M_{\odot}$ to disregard halos with fewer than 100 particles because smaller halos are unreliable and dependent on which halo finder we use. The central is assigned to the centre of mass of the halo with probability $\langle n_{\rm cen}\rangle$. The satellites are assigned to each particle within the halo with probability $p = \langle n_{\rm sat}\rangle/n_{\rm particle}$, where $n_{\rm particle}$ is the total number of particles within the halo. We assign satellite galaxies at particle positions instead of by fitting a smooth profile (e.g.~Navarro-Frenk-White profile; \citealt{1996Navarro, 1997Navarro}) since we want to allow satellites to trace halo substructure and avoid assumptions about the isotropy or equilibrium state of the halos. 

Looping through all the dark matter particles and assigning galaxies with a small probability introduces significant shot noise, which leads to uncertainties in our $\bpg$ and $\qeff$ signals not greater than $10\%$ (refer to Figure~\ref{fig:results_real} and Figure~\ref{fig:results_rsd}). For most of our analysis, the $\bpg$ and $\qeff$ signals are strong enough that this shot noise is not important. However, in some cases we require much higher signal-to-noise, such as in Section~\ref{sec:hod_depend}, where we investigate percent level perturbations to the $\bpg$ and $\qeff$ signal due to variations in HOD parameters. To suppress shot noise, we seed the Python pseudo random number generator with the same number at the start of generating the mock galaxy catalogue for each simulation. Thus, when we loop through all the dark matter particles to populate them with galaxies, each dark matter particle is called with the same pseudo random number on each run. This way, not only do we get reproducible results when we re-run the same HOD, but also small changes in the HOD make small changes in the galaxy distribution, thereby sharply reducing the noise in the numerical derivatives of the correlation functions with respect to HOD parameters.  

To further suppress the shot noise, we populate each of the 16 simulation boxes with the same HOD 16 different times, each time with a different seed. Then we take the average $\qeff$ of the $16^2$ realizations. Note that the reseeding is only done for Figure~\ref{fig:results_rsd_zoom} and Figure~\ref{fig:dQeff}, due to computational limit. 

Once we have compiled the mock LRG catalogue from our HOD, the LRG pairs are generated with a pair search on a kd-tree of LRG locations. For our simulation boxes, this HOD typically generates around $5.3\times 10^5$ LRGs per box. For maximum pair separation $d_{p, \rm{max}} = 10$~Mpc in real space, we generate around $8.7\times 10^5$ pairs per box, of which around $2.9\times 10^5$ are identified as 1-halo pairs, i.e. the two LRGs forming the pair occupy the same halo. 
We clarify that all pairs within maximum pair separation $d_{p, \rm{max}}$ are counted in our pair catalogues, e.g. we include all 3 pairs generated by a close triplet. Also for the purpose of this paper, we do not consider effects of fiber collisions or survey boundaries. 

\begin{figure}
\centering
 \hspace*{-0.4cm}
\includegraphics[width=3.6in]{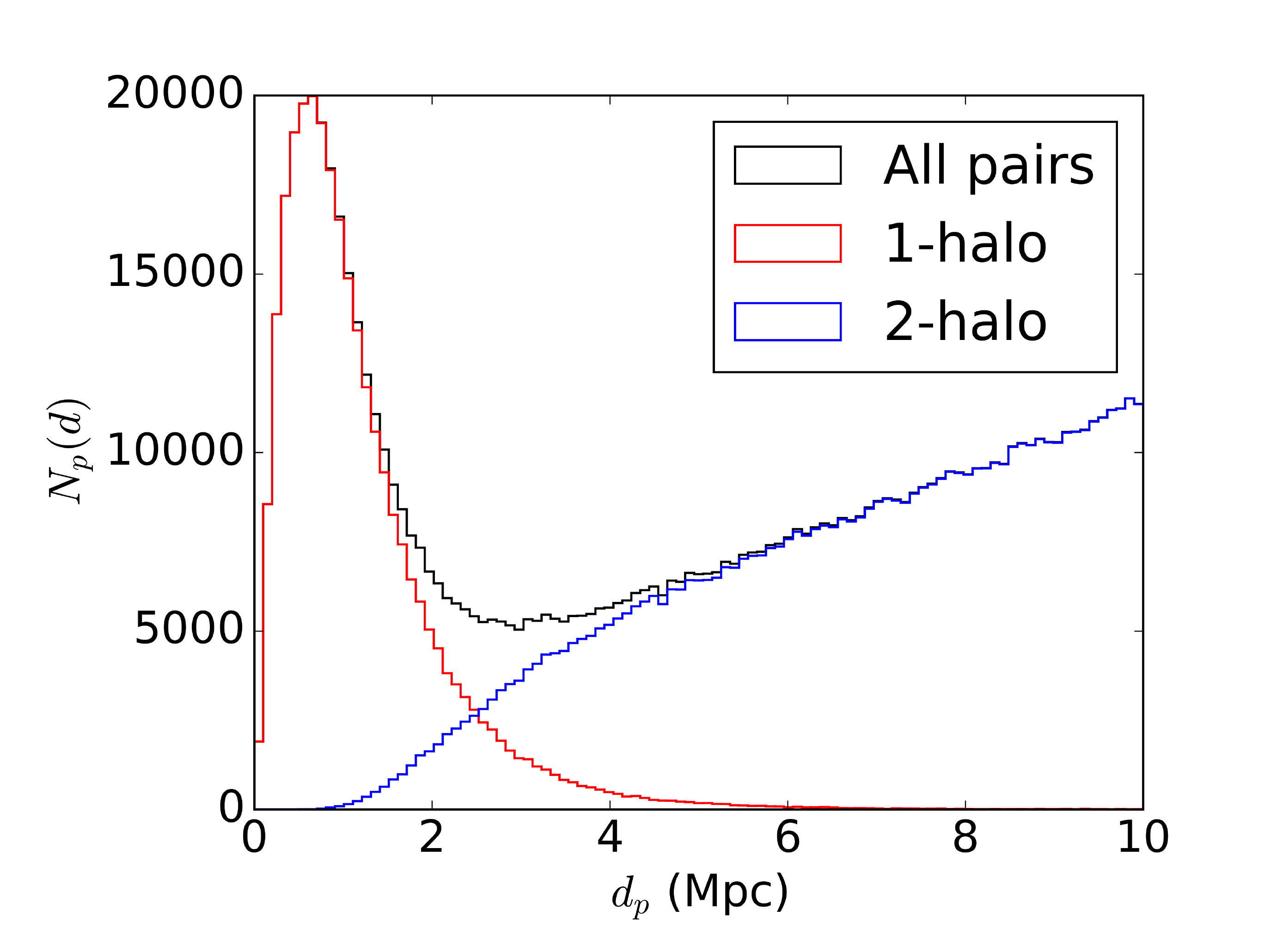}
\vspace{-0.6cm}

\caption{The real-space pair separation distribution up to $d_{p, \rm{max}} = 10$~Mpc for one of the 16 simulation boxes. The red histogram represents the 1-halo pairs, whereas the blue histogram represents the 2-halo pairs, and the black histogram is the sum of both. }
\label{fig:hist_dp}
\end{figure}

Figure~\ref{fig:hist_dp} shows the distribution of pair separation in all 16 simulation boxes in real space. We see that at $d_p < 2$~Mpc, 1-halo pairs dominate, while at $d_p > 2$~Mpc, 2-halo pairs dominate. 

\begin{figure}
\centering
 \hspace*{-0.4cm}
\includegraphics[width=3.6in]{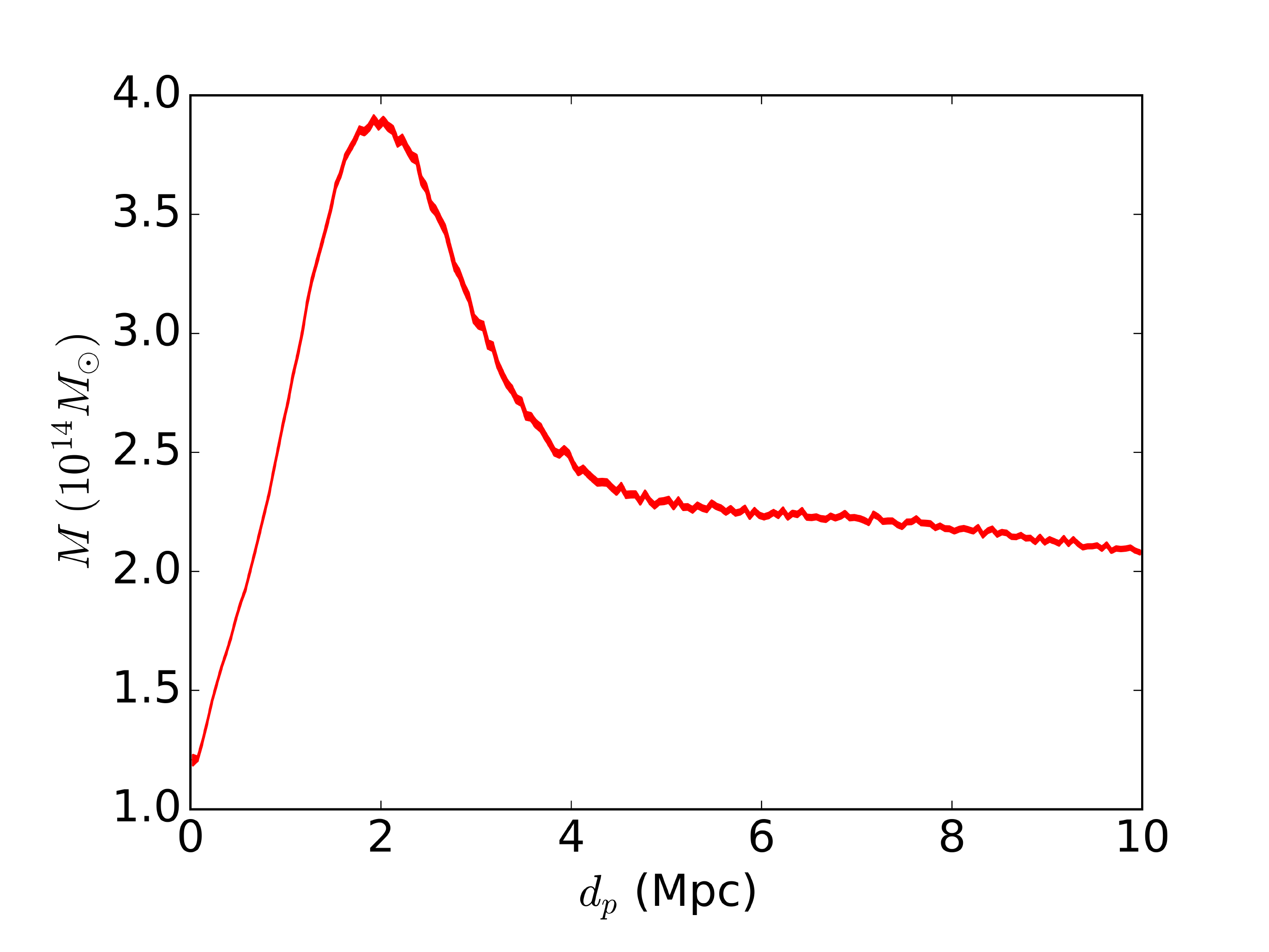}
\vspace{-0.6cm}

\caption{The halo mass $M$ of the pairs as a function of real-space pair separation $d_p$ in real space. For 1-halo pairs, $M$ is simply the mass of the host halo, while for 2-halo pairs, $M$ is the sum of the mass of the two halos. The values plotted are averaged across all 16 simulations. The linewidth shows the sample standard deviation at that $d_p$. We see that the halo mass strongly peaks at around 2~Mpc, where we have mostly 1-halo pairs but also a significant fraction of 2-halo pairs. }
\label{fig:mhalo_dp}
\end{figure}

In Figure~\ref{fig:mhalo_dp}, we plot the halo mass corresponding to the LRG pair as a function of pair separation $d_p$ in real space. For 1-halo pairs, the halo mass $M$ is simply the mass of the host halo, and for 2-halo pairs, $M$ is the sum of the mass of the two halos. The width of the shaded region shown on the plot is twice the sample standard deviation at that $d_p$. We see a distinct peak at around $d_p \approx 2$~Mpc, suggesting that close LRG pairs at approximately 2~Mpc apart sample the most massive halos, whereas at separation less or greater than $d_p \sim 2$~Mpc, the halo mass of the pair decreases.

\begin{figure}
\centering
 \hspace*{-0.4cm}
\includegraphics[width=3.6in]{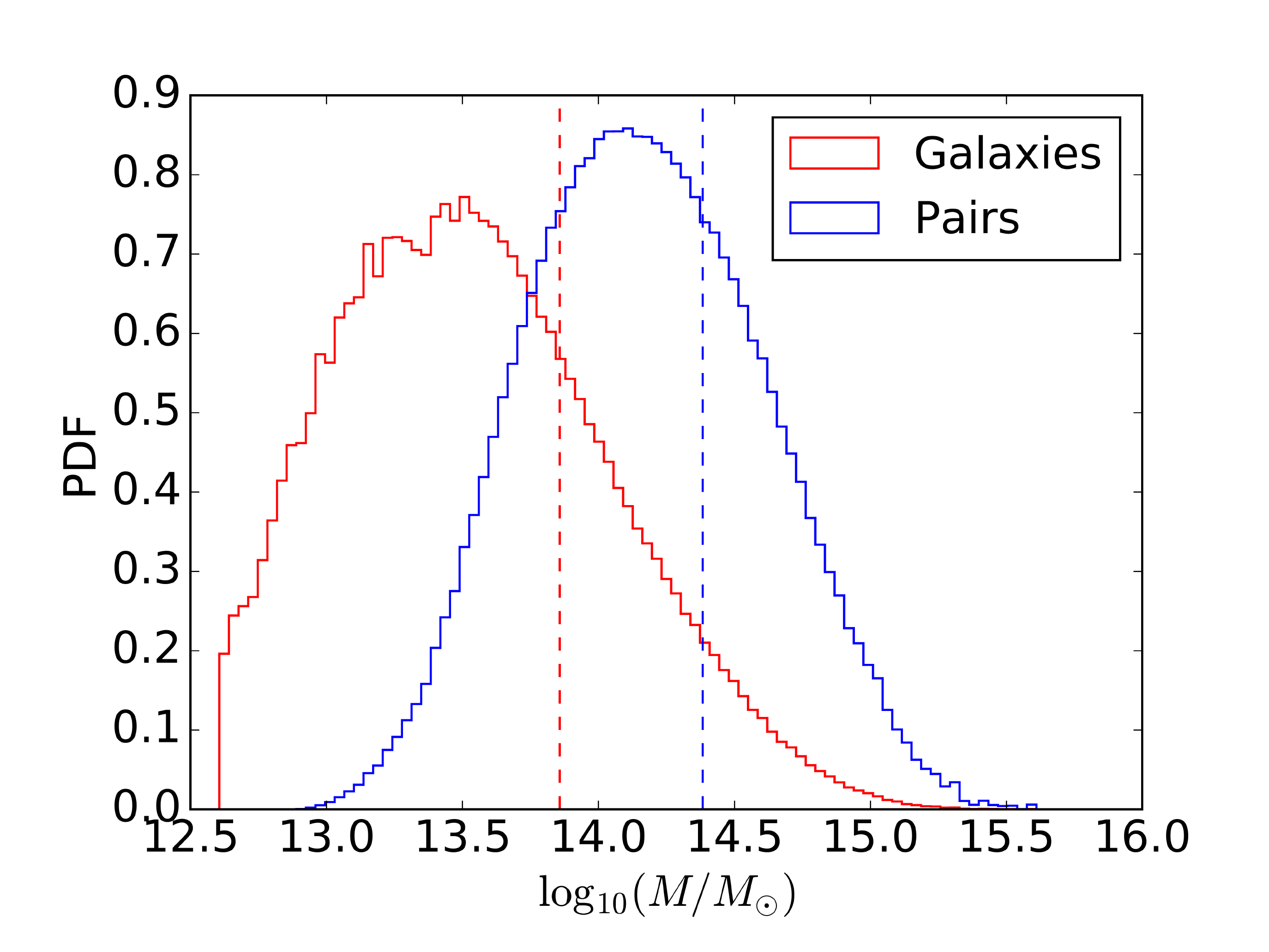}
\vspace{-0.6cm}

\caption{The halo mass $M$ probability density distribution for galaxies (in red) and for pairs (in blue). The dashed curves showcase the average halo mass of the galaxies and the pairs. The pair sample used for this plot has a maximum pair separation $d_{p, \rm{max}} = 10$~Mpc in real space. We see that the pairs at this separation trace more massive halos than galaxies alone. }
\label{fig:mhalo_hist}
\end{figure}

Figure~\ref{fig:mhalo_hist} shows the halo mass probability density distribution for galaxies (in red) and for pairs (in blue), with the average halo mass denoted by the dashed curves. The pair sample shown on this figure has a maximum pair separation $d_{p, \rm{max}} = 10$~Mpc in real space. We see that the galaxies sample halos of approximately $10^{13.5} M_{\odot}$ whereas the pairs sample more massive halos at around $10^{14.2} M_{\odot}$. This plot motivates the point in Section~\ref{sec:intro} that close pairs are biased tracers of more massive halos and that statistics on the pair field can be used to constrain the high mass regime of the HOD.

\subsection{2PCF}
\label{sec:2PCF}

By the pair-counting definition of 2PCF, we can calculate the 2PCF as 
\begin{equation}
\xi(d_p) = \frac{2 N_p(d_p)}{\bar{n}_g N_g\Delta V_{\rm{shell}}} - 1,
\label{equ:2PCF_d_collapsed}
\end{equation}
where $N_p(d_p)$ is the number of pairs with separation $d_p$ and $\Delta V_{\rm{shell}} = 4\pi d_p^2 \Delta d_p$ is the shell volume. $\bar{n}_g$ denotes the LRG number density, and $N_g$ denotes the total number of LRGs in the sample. 

If we consider RSD, the LOS separation of the pair $d_{\parallel}$ becomes largely meaningless due to velocity dispersion along the LOS. 
Thus, we calculate the LOS projected 2PCF, $w(d_{\perp})$, as a function of the perpendicular (to LOS) component of pair separation $d_{\perp}$,
\begin{align}
w(d_{\perp}) & = \int_{-d_{\parallel, \rm{max}}}^{d_{\parallel, \rm{max}}} \xi(d_{\parallel}, d_{\perp}) d(d_{\parallel}) \nonumber \\
& = 2 d_{\parallel, \rm{max}}\left( \frac{2 \sigma_p(d_{\perp})}{\bar{n}_g N_g\Delta V_{\perp}} - 1\right),
\label{equ:2PCF_los_integrated_explicit}
\end{align}
where $d_{\parallel, \rm{max}} = \sqrt{d_{p, \rm{max}}^2 - d_{\perp}^2}$ is the maximum pair separation along the LOS given a maximum pair separation $d_{p,\rm{max}}$. 
$\sigma_p(d_{\perp})$ is the integrated column pair density, and $\Delta V_{\perp} = 2\pi \Delta d_{\perp} \times 2 d_{\parallel, \rm{max}}$. Unlike the real space 2PCF given by equation~\ref{equ:2PCF_d_collapsed}, the LOS projected 2PCF can be inferred from observed LRG distributions without redshift information. For this reason, we use projected 2PCF instead of the full 2PCF in Section~\ref{sec:2pf_hod} as well. 

\begin{figure}
\centering
 \hspace*{-0.7cm}
\includegraphics[width=3.9in]{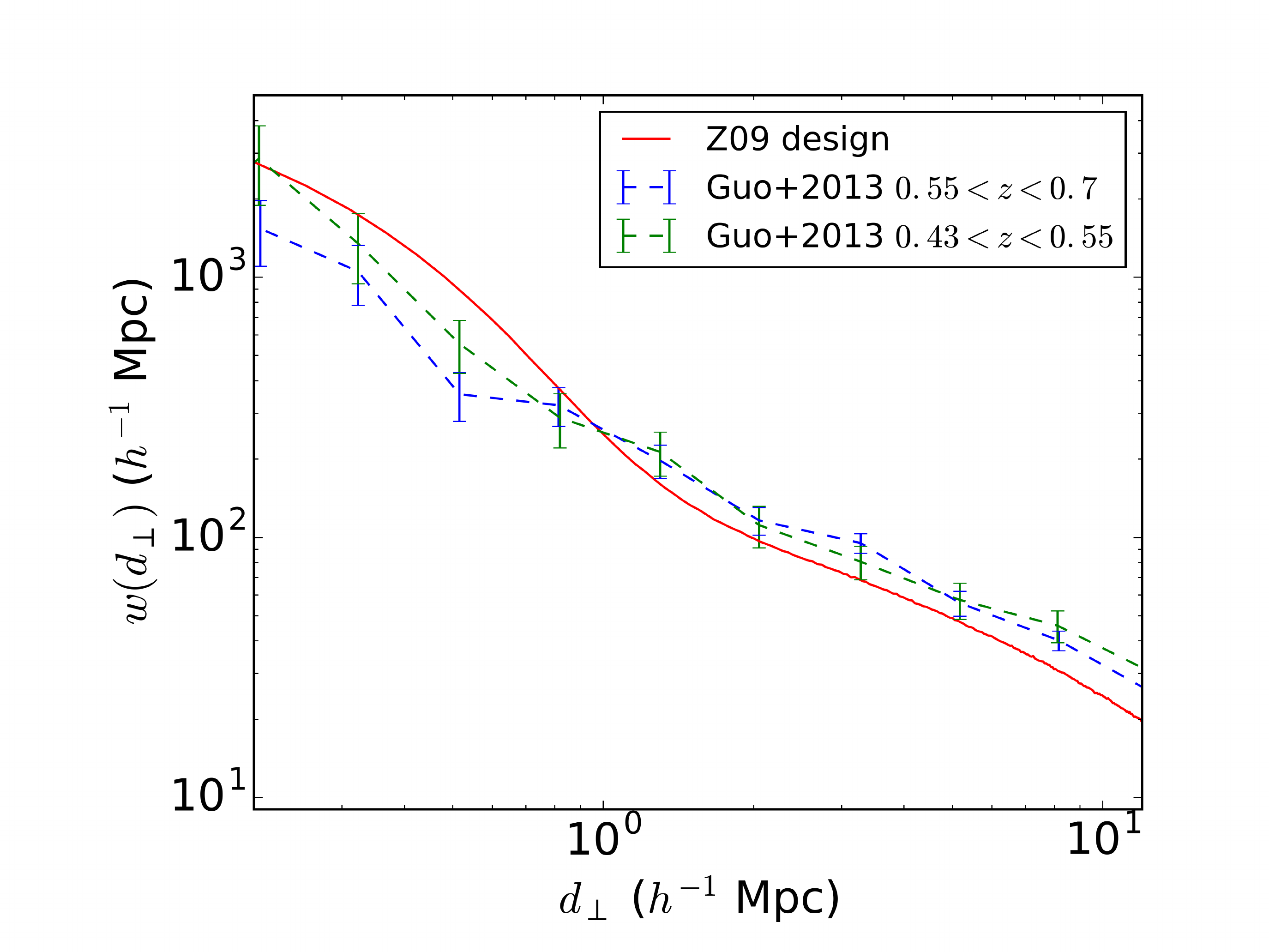}
\vspace{-0.6cm}
\caption{The real space 2PCF as a function of pair separation generated from the simulation box is in red solid line. The 2PCFs from SDSS CMASS sample in two different redshift bins are plotted in dashed blue line and dashed green line respectively.}
\label{fig:2PCF_observ}
\end{figure}

 Figure~\ref{fig:2PCF_observ} shows $w(d_{\perp})$ calculated from our LRG pair catalogue. We also plot in dashed blue and green curves the projected 2PCFs from the CMASS luminous red galaxies sample of Sloan Digital Sky Survey III (SDSS-III; \citealt{2011Eisenstein, 2013Dawson}) within the redshift range $0.55 < z < 0.7$ and $0.43 < z < 0.55$, respectively. These 2PCFs are directly adapted from \citet{2013Guo}. Our 2PCF is roughly consistent with the observed 2PCFs from CMASS red galaxies. The inconsistency in shape can be attributed to a combination of differences in cosmology, problems with halo finders, and oversimplifications in our HOD model itself. However, for the purpose of this methodology paper, the exact form of the HOD is not essential, thus a detailed fit to the observed 2PCF is not important for our analysis. Applications of our methodology to more sophisticated halo occupation models are deferred to future papers. 

\subsection{Mass assignment function}
\label{sec:tsc}

To be able to utilize FFTs to efficiently compute the power spectra of the LRG and LRG pair fields, we require a mass assignment function that assigns LRGs (and pairs) onto a regular 3D grid in such a way that generates a smooth LRG (and pair) density field. For the purpose of this paper, we assign LRGs and pairs onto an evenly spaced $512^3$ grid spanning the full simulation box using the Triangular Shaped Cloud (TSC) method \citep{2010Jeong}. 
For a simulation box size $L_{\rm{box}} = 1100 h^{-1}$~Mpc with 512$^3$, the cell size is $\Delta x \approx 2.15 h^{-1}$~Mpc. Thus, a TSC of base length $2\Delta x$ spans 27 cells on the grid, with 3 cells along each dimension. The fact that each TSC can span at most 3 cells along each direction becomes important in the next section as we design our custom weighting functions. 

\subsection{Weighting function}
\label{sec:weighting}

In Section~\ref{sec:method}, we present our method of efficiently computing the cross-correlation between galaxy pairs and galaxies, from which we infer the squeezed 3PCF. This routine requires a robust weighting function for the pair-galaxy cross-correlation that satisfies the following three criteria.

(a) The weighting function should prevent self-counting, where one of the two galaxies within the pair is considered as the third galaxy when performing the cross-correlation. This means that the weighting function should have a hollow spherical cavity in the centre. The radius of the cavity should be $r_{\rm{cut}} \approx  d_{p, \rm{max}}/2 + 2\Delta x$, where $d_{p, \rm{max}}$ is the maximum pair separation and $\Delta x$ is the cell size. The extra $2\Delta x$ comes from avoiding the TSC of the pair, which can extend a maximum of 2 cells outwards from either galaxy along the axis of the pair. 

(b) The weighting function should minimize the FoG effect when we apply our methodology to redshift space data, where the correlation along the LOS becomes largely meaningless. To disregard correlation between pairs and galaxies along the LOS, we  ``drill" a cylindrical hole through the centre of the weighting function parallel to the LOS. The radius of the cylindrical hole should be $r_{\rm{cut}} \approx  d_{p, \rm{max}}/2 + 2\Delta x$ for the same reason as explained in (a). 

(c) The weighting function should sample scales approximately in the range $k = 0.01\sim 0.1 h$~Mpc$^{-1}$. The scale range is limited on the large scale by the simulation box size, where we cannot meaningfully interpret scales larger than the box size, which translates to a minimum frequency of $k_{\rm{min}} = 2\pi/L_{\rm{box}} \approx 0.006 h$~Mpc$^{-1}$. The scale range is limited on the small scale by self-counting, so we do not want the third galaxy to be too close to the pair that their TSCs overlap. We also avoid small scale contribution since we want to be computing bias in the linear regime. 

\begin{figure}
\centering
 \hspace*{-0.4cm}
\includegraphics[width=3.6in]{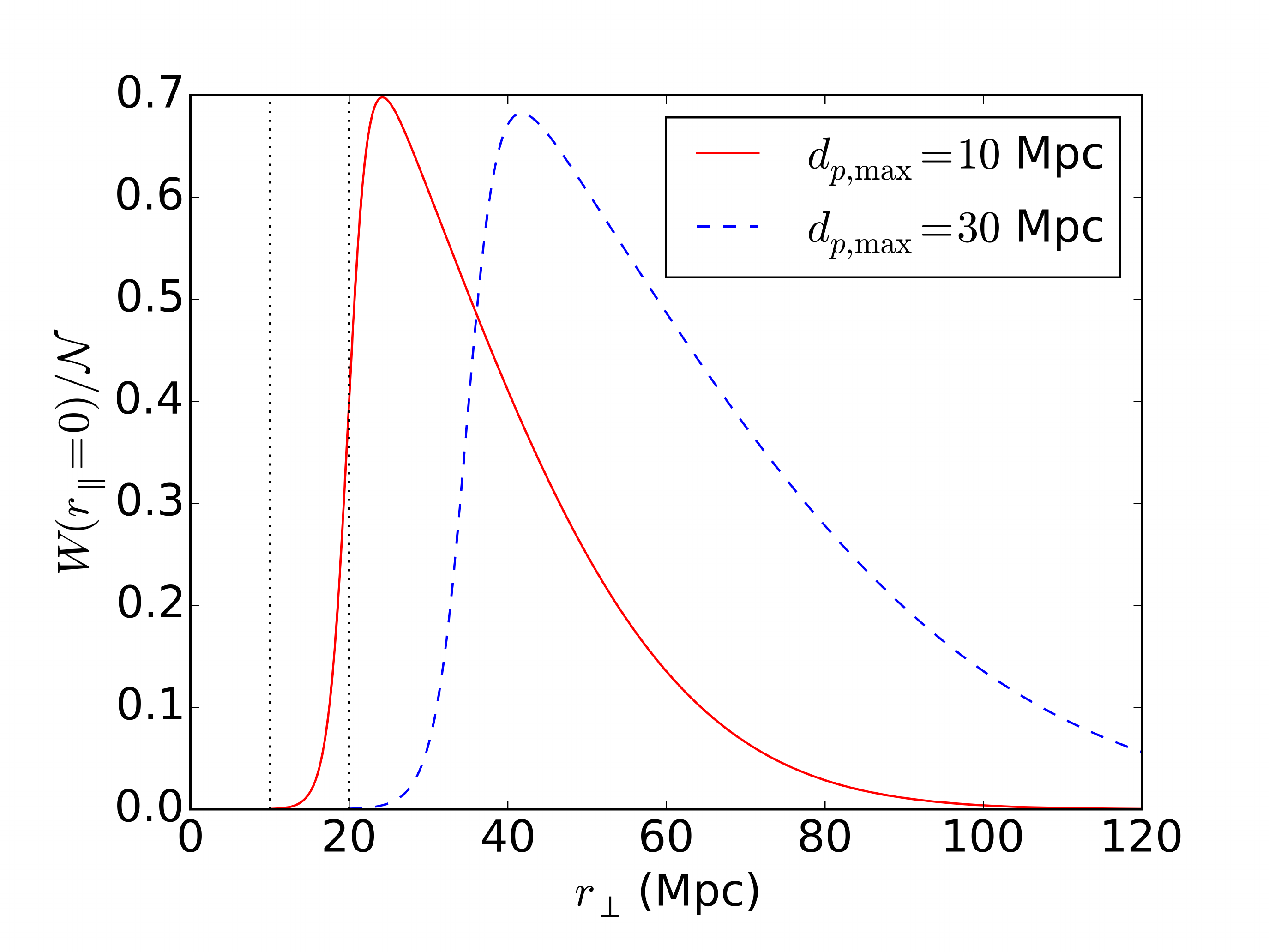}
\vspace{-0.5cm}
\caption{The weighting functions for $d_{p, \rm{max}} = 10$~Mpc (red solid line) and $d_{p, \rm{max}} = 30$~Mpc (blue dashed line) as defined in equation~\ref{equ:weighting}. We only show a 2D cutaway of the 3D function at $r_{\parallel} = 0$. The dotted black lines mark $r_{\rm{cut}} = 10$~Mpc and $r_{\rm{cut}} = 20$~Mpc.}
\label{fig:weightings}
\end{figure}

In addition to these requirements, we also want the weighting function to be mostly smooth and analytic for ease of performing FFTs and to avoid high $k$ modes in the weighting function. Specifically, our weighting function is given in real space by,
\begin{align}
& W(r_{\perp}, r_{\parallel}) = \nonumber \\ 
& \begin{cases}
\mathcal{N}\frac{1}{1+\exp\left[-s_0(r_{\perp} - r_s)\right]}\exp\left[-\frac{1}{2}\left(\frac{r_{\perp}^2}{\sigma^2}+\frac{r_{\parallel}^2}{\sigma^2}\right)\right], & r_{\perp} \geq r_{\rm{cut}}\\
0, & r_{\perp} < r_{\rm{cut}}
\end{cases}
\label{equ:weighting}
\end{align}
where $\mathcal{N}$ is a normalization constant. $r_{\perp}$ and $r_{\parallel}$ are distance between the third galaxy and the centre of the pair, projected perpendicular and parallel to the LOS respectively. The function is parameterized by $\sigma$, $r_s$, $s_0$ and $r_{\rm{cut}}$, which are determined by the maximum pair separtion$d_{p, \rm{max}}$ that we choose. Although the function is divided into two segment and thus not fully analytical, such engineering is necessary to remove contamination from pair self counting. We choose the parameters such that the jump in $W(r_{\perp}, r_{\parallel})$ at $r_{\rm{cut}}$ is small (refer to Figure~\ref{fig:weightings}). 

For the rest of the analysis in this paper, we use $d_{p, \rm{max}} = 10$~Mpc when working in real space and $d_{p, \rm{max}} = 30$~Mpc when working in redshift space. We choose $d_{p, \rm{max}} = 30$~Mpc in redshift space because the typical velocity dispersion along the LOS translates to $\sim 10$~Mpc displacement, so we extend the pair search in both direction by $10$~Mpc to capture most of the pairs. The parameters of the weighting function for both cases are summarized in Table~\ref{tab:weightings}. 
\begin{table*}
\centering
\scalebox{1}{
\begin{tabular}{ c | c c c c c c}
\hhline {=======}
$d_{p, \rm{max}}$ (Mpc) & $\sigma$ (Mpc) & $r_s$ (Mpc) & $s_0$ (Mpc$^{-1}$) & $r_{\rm{cut}}$ (Mpc) & $d_w$ (Mpc) & $\xi(d_w)$ \\ 
\hline
10 & 30 & 20 & 0.8 & 10 & 30.9 & 0.362 \\ 

30 & 50 & 35 & 0.5 & 20 & 52.6 & 0.104 \\ 
\hline 
\end{tabular} 
}
\caption{Parameters of the weighting functions for $d_{p, \rm{max}} = 10$~Mpc and $d_{p, \rm{max}} = 30$~Mpc. Again the cutoff is given by $r_{\rm{cut}} \approx  d_{p, \rm{max}}/2 + 2\Delta x$.}
\label{tab:weightings}
\end{table*}

Figure~\ref{fig:weightings} shows a 2D cutaway of the weighting functions at $r_{\parallel} = 0$ in the positive half of $r_{\perp}$. The solid red curve shows the weighting function for $d_{p, \rm{max}} = 10$~Mpc and the dashed blue curve shows the weighting function for $d_{p, \rm{max}} = 30$~Mpc. The dotted black lines denote the location of $r_{\rm{cut}}$ to show that the jump at $r_{\rm{cut}}$ is small.

\begin{figure}
\centering
 \hspace*{-0.4cm}
\includegraphics[width=3.8in]{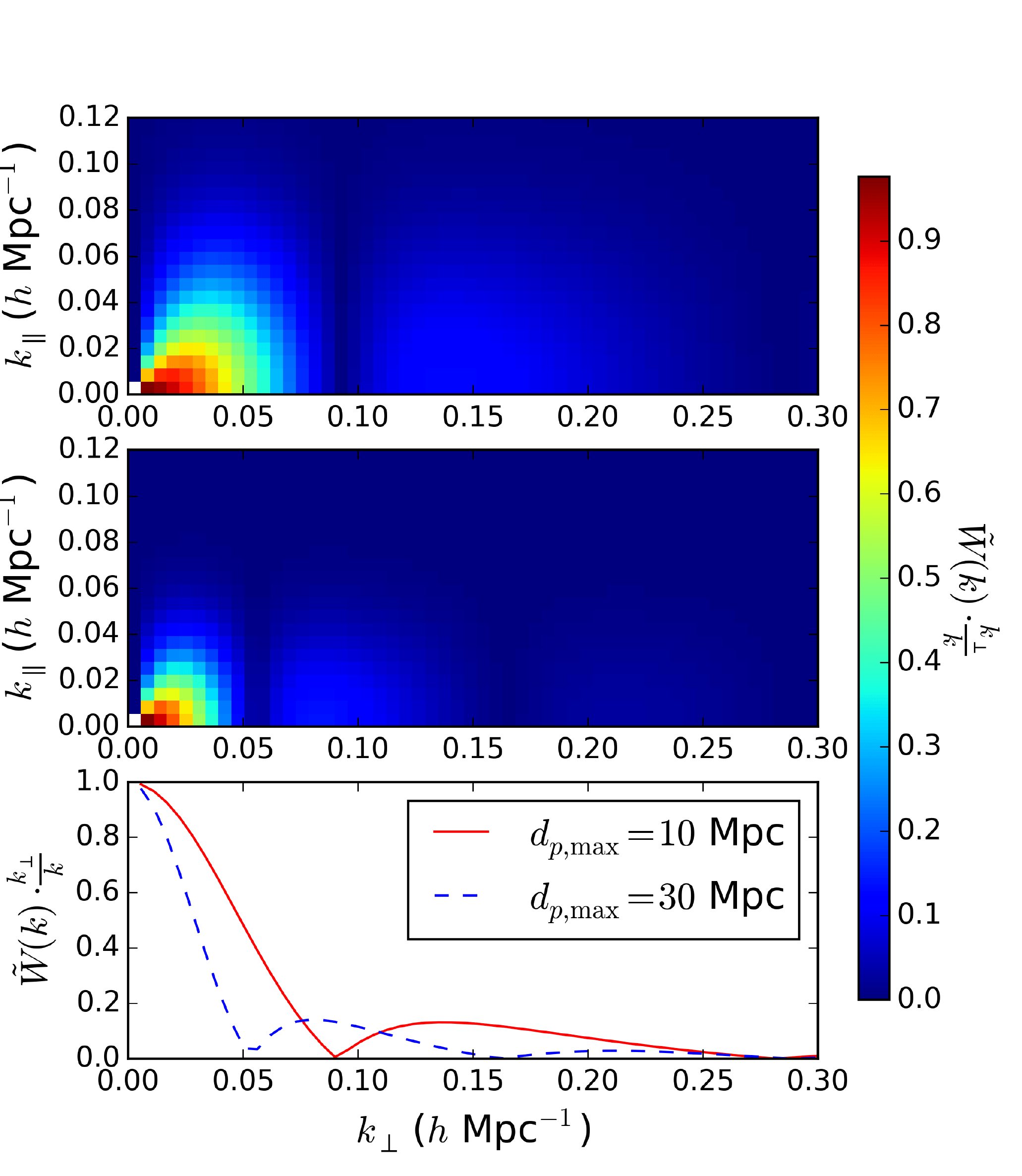}
\vspace{-0.5cm}
\caption{The Fourier Transforms of the weighting function for $d_{p, \rm{max}} = 10$~Mpc (top panel) and $d_{p, \rm{max}} = 30$~Mpc (middle panel). The bottom panel shows cutaways of $\tilde{W}(k)\ \frac{k_{\perp}}{k}$ at $k_{\parallel } = 0$ for the $d_{p, \rm{max}} = 10$~Mpc case (red solid curve) and the $d_{p, \rm{max}} = 30$~Mpc case (blue dashed curve).}
\label{fig:weightings_fft}
\end{figure}

According to equation~\ref{equ:bpg_final}, the weighting function is multiplied with the pair density field in Fourier space as a sampling weight for different $\textbf{k}$ modes. The Fourier Transforms of the weighting functions are plotted in Figure~\ref{fig:weightings_fft}, where the top panel shows the weighting function for $d_{p, \rm{max}} = 10$~Mpc whereas the middle panel shows the weighting function for $d_{p, \rm{max}} = 30$~Mpc. The bottom panel shows cutaways of the two cases at $k_{\parallel} = 0$. The colorbar gives $\tilde{W}\  k_{\perp}/k$, which is proportional to the number of modes sampled by the weighting function in Fourier space. We see that both weighting functions are indeed heavily sampling scales between $0.01\sim 0.1 h$~Mpc$^{-1}$. The first and second order ringing seen at higher frequency are due to the steep drop-off in the real space weighting function towards small $r_{\perp}$. However, the ringings have much lower integrated weight in $k$-space and are not expected to affect our analysis significantly. 

Now that we have defined the weighting functions, we can calculate $d_w$ (equation~\ref{equ:dw}) and $\xi(d_w)$ (equation~\ref{equ:xi_dw}), which we need to compute the squeezed 3PCF coefficient $\qeff$ from the pair-galaxy bias $b_{p,g}$.
To do so, we employ the 2PCF given by Figure~2 and Figure~3 in \citet{2005Eisenstein}, which is constructed from the linear power spectra calculated using CMBFAST \citep{1996Seljak, 1998Zaldarriaga, 2000Zaldarriaga}. For any of our weighting functions, we fit the 2PCF with a power law in the scale range sampled by the weighting function. Then we use the best fit exponent $p$ to calculate $d_w$ and $\xi(d_w)$. Specifically, the $d_w$ and $\xi(d_w)$ for our $d_{p, \rm{max}} = 10$~Mpc and 30~Mpc are given in the last two columns of Table~\ref{tab:weightings}. It should be pointed out that the analysis in \citet{2005Eisenstein} uses a slightly different cosmology, where $h = 0.7$, $\Omega_m h^2 = 0.13$ and $\Omega_b h^2 = 0.024$. However, the correction is small and does not meaningfully change our calculations. 

Now that we have defined our mass assignment function and our weighting functions, we apply the TSC to the LRGs and LRG pairs in all 16 simulation boxes to generate smooth density fields of the LRGs and pairs on our grid. The resulting density fields are then convolved with the appropriate weighting function to compute the pair-galaxy bias $\bpg$ (equation~\ref{equ:bpg_final}) and the reduced 3PCF $\qeff$ (equation~\ref{equ:Qeff_notsq}). 

\section{Results and Discussion}
\label{sec:results}

In this section, we present the results of implementing our efficient method of calculating $\bpg$ and $\qeff$ using the N-body simulation and the Z09 HOD. The results shown are averaged over the 16 cosmological boxes to reduce noise. Section~\ref{sec:results_real} presents the results in real space, and Section~\ref{sec:results_rsd} presents the results in redshift space.

\subsection{Results in real space}
\label{sec:results_real}

\begin{figure}
\centering
 \hspace*{-0.4cm}
\includegraphics[width=3.6in]{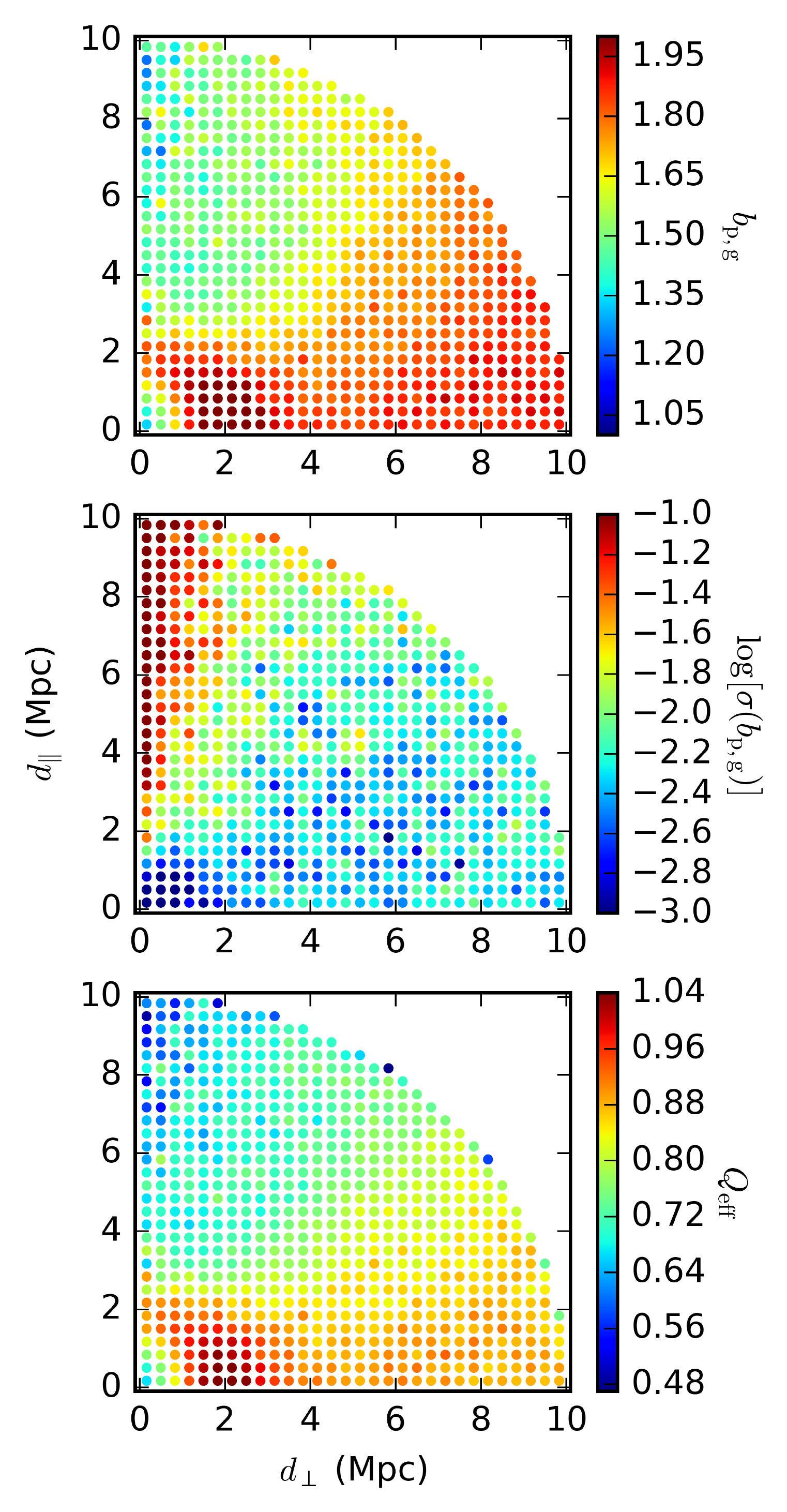}
\vspace{-0.5cm}
\caption{$\bpg$ and $\qeff$ results in real space. The top panel plots the pair-galaxy bias distribution as a function of real-space pair separation. The middle panel plots the corresponding uncertainty of the pair-galaxy bias distribution. The uncertainty is calculated as the standard deviation of the pair-galaxy bias calculated for all 16 simulation boxes. The bottom panel plots the corresponding reduced 3PCF $\qeff$.
The weighting function is the Gaussian-sigmoid function for $d_{p, \rm{max}} = 10$~Mpc as described in Section.~\ref{sec:weighting}.}
\label{fig:results_real}
\end{figure}

We average the $\bpg$ calculated from all 16 simulation boxes at $z = 0.5$ and plot the averaged real space $\bpg$ in the top panel of Figure~\ref{fig:results_real}. We have chosen maximum pair separation $d_{p, \rm{max}} = 10$~Mpc and a Gauss-Sigmoid weighting function with the following parameters: $\sigma = 50$~Mpc, $r_s = 30$~Mpc, $s_0 = 0.5$~Mpc$^{-1}$ and $r_{\rm{cut}} = 15$~Mpc (Refer to Section~\ref{sec:weighting} for details). The associated uncertainty map is given in middle panel. The bottom panel shows the reduced 3PCF $Q_{\rm eff}$ in real space, as calculated from $\bpg$ using equation~\ref{equ:Qeff_notsq}. 

The top panel of Figure~\ref{fig:results_real} shows that the pair-galaxy bias is close to 1.3 at pair separation less than $\sim 1$~Mpc, but dramatically increases by a factor of 2 and peaks at $\sim 2$~Mpc, before decreasing again towards greater pair separation. The bottom panel shows that the squeezed 3PCF coefficient $\qeff$ recovers the same behavior. We can analytically estimate the value of $\bpg$ by computing the halo bias as a function of halo mass and then take the ratio of the halo bias averaged over all pairs and the halo bias averaged over all galaxies (using the halo mass function of the pairs and galaxies shown in Figure~\ref{fig:mhalo_hist}). Our estimate yields $\bpg \approx 1.7$, which is consistent with the values we see in the top panel of Figure~\ref{fig:results_real}.

Referring to Figure~\ref{fig:hist_dp} and Figure~\ref{fig:mhalo_dp}, we see that pairs with separation $< 1$~Mpc are mostly 1-halo pairs with lower halo mass. $\bpg$ is approximately 1.3 at $d_p < 1$~Mpc. The fact that the pair-galaxy bias $\bpg$ is greater than 1 shows that close galaxy pairs are more biased than individual galaxies. 
Pairs with separation $\sim 2$~Mpc consist of $\sim 70\%$ larger 1-halo pairs and $\sim 30\%$ 2-halo pairs. Figure~\ref{fig:mhalo_dp} shows that these pairs trace the most massive halos, thus giving them the highest bias. 

Figure~\ref{fig:results_real} also shows a strong anisotropic signal in $\bpg$ and $\qeff$ at large pair separation $d_p > 4$~Mpc. The anisotropy is due to the fact that the weighting function samples favorably volume in the transverse direction and avoids volumes along the LOS direction. When a pair is oriented transverse to the LOS, it establishes the transverse direction as the preferred direction for the large-scale filament. Thus, when a pair is perpendicular to the LOS, the weighting function overlaps the most with the large-scale filament, giving rise to much stronger clustering. Conversely, when a pair is oriented along the direction of LOS, the filament is preferentially aligned with the LOS, which does not overlap significantly with the weighting function, thus giving rise to weaker clustering. The large $\bpg$ value at $d_p > 4$~Mpc suggests that the filament, not just the halo, is also strongly biased against the galaxy field.

The middle panel of Figure~\ref{fig:results_real} shows that most of the $b_{p,g}$ values have percent level uncertainties, except the ones towards the top left corner of the plot, where the small sample size contributes to the high noise. 

 The bottom panel of Figure~\ref{fig:results_real} shows that the reduced 3PCF recovers the same behavior as seen in the pair-galaxy bias $b_{p,g}$. This is consistent with our conclusion that in the strictly squeezed limit, the 3PCF would basically reduce to $Q_{\rm{eff}} = b_{p,g}/2$. At pair separation $d_p > 5$~Mpc, we continue to see the anisotropy in bias due to the filamentary structure aligning with the preferred direction of the weighting function. However, we also see suppression of a few percent in $Q_{\rm eff}$ due to the $\xi(d_w)$ term. 
The associated uncertainty map of $\qeff$ is omitted, but shares the same properties as the uncertainty in $\bpg$, but with approximately half the amplitude.

\subsection{Results in redshift space}
\label{sec:results_rsd}

\begin{figure}
\centering
 \hspace*{-0.4cm}
\includegraphics[width=3.6in]{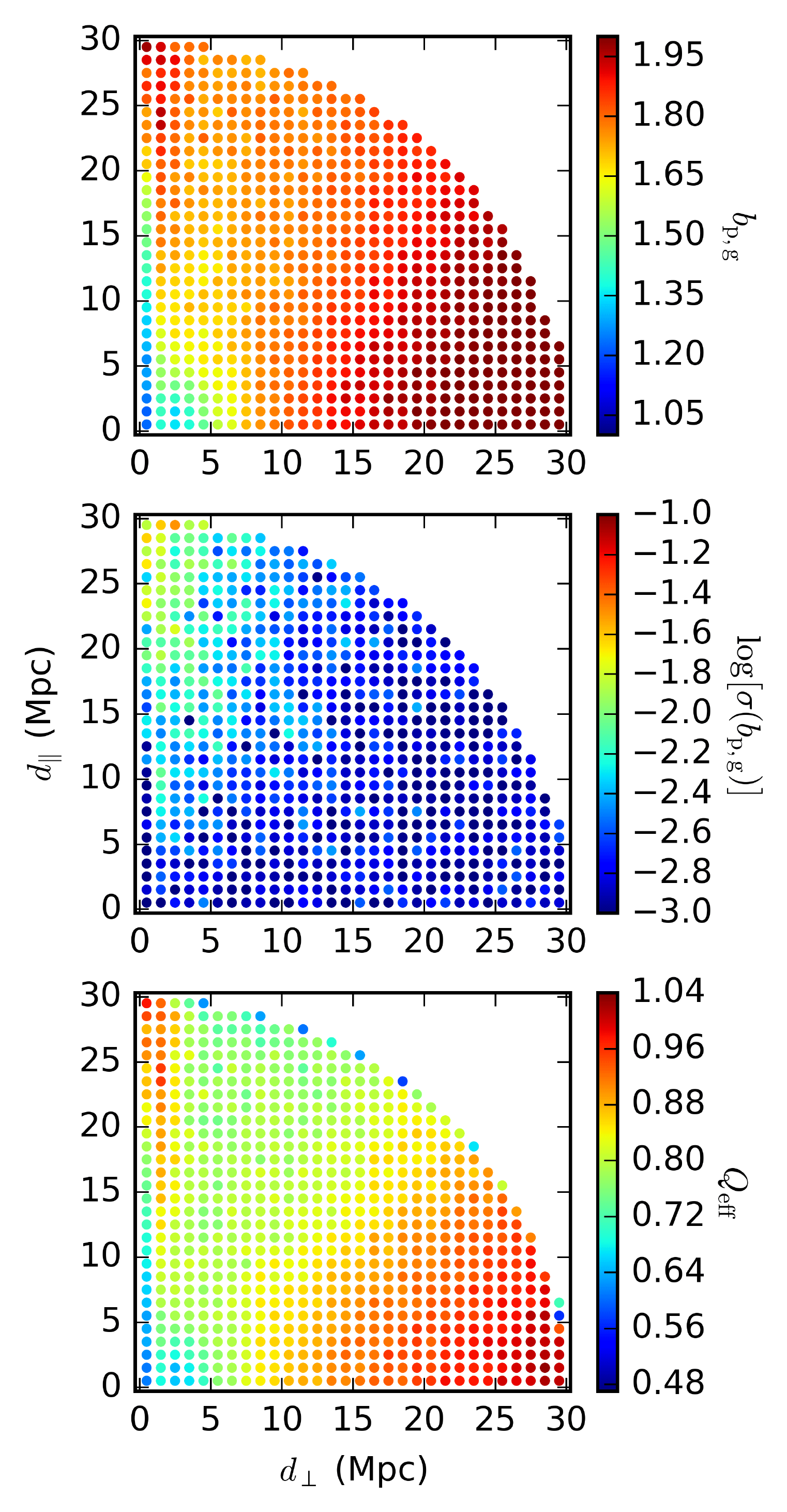}
\vspace{-0.5cm}
\caption{$\bpg$ and $\qeff$ results in redshift space. The top panel plots the pair-galaxy bias distribution as a function of redshift-space pair separation. The middle panel plots the corresponding uncertainty of the pair-galaxy bias distribution. The uncertainty is calculated as the standard deviation of the pair-galaxy bias calculated for all 16 simulation boxes. The bottom panel plots the corresponding reduced 3PCF $\qeff$. 
The weighting function is the Gaussian-sigmoid function for $d_{p, \rm{max}} = 30$~Mpc as described in Section.~\ref{sec:weighting}.}
\label{fig:results_rsd}
\end{figure}

\begin{figure}
\centering
 \hspace*{-0.6cm}
\includegraphics[width=3.8in]{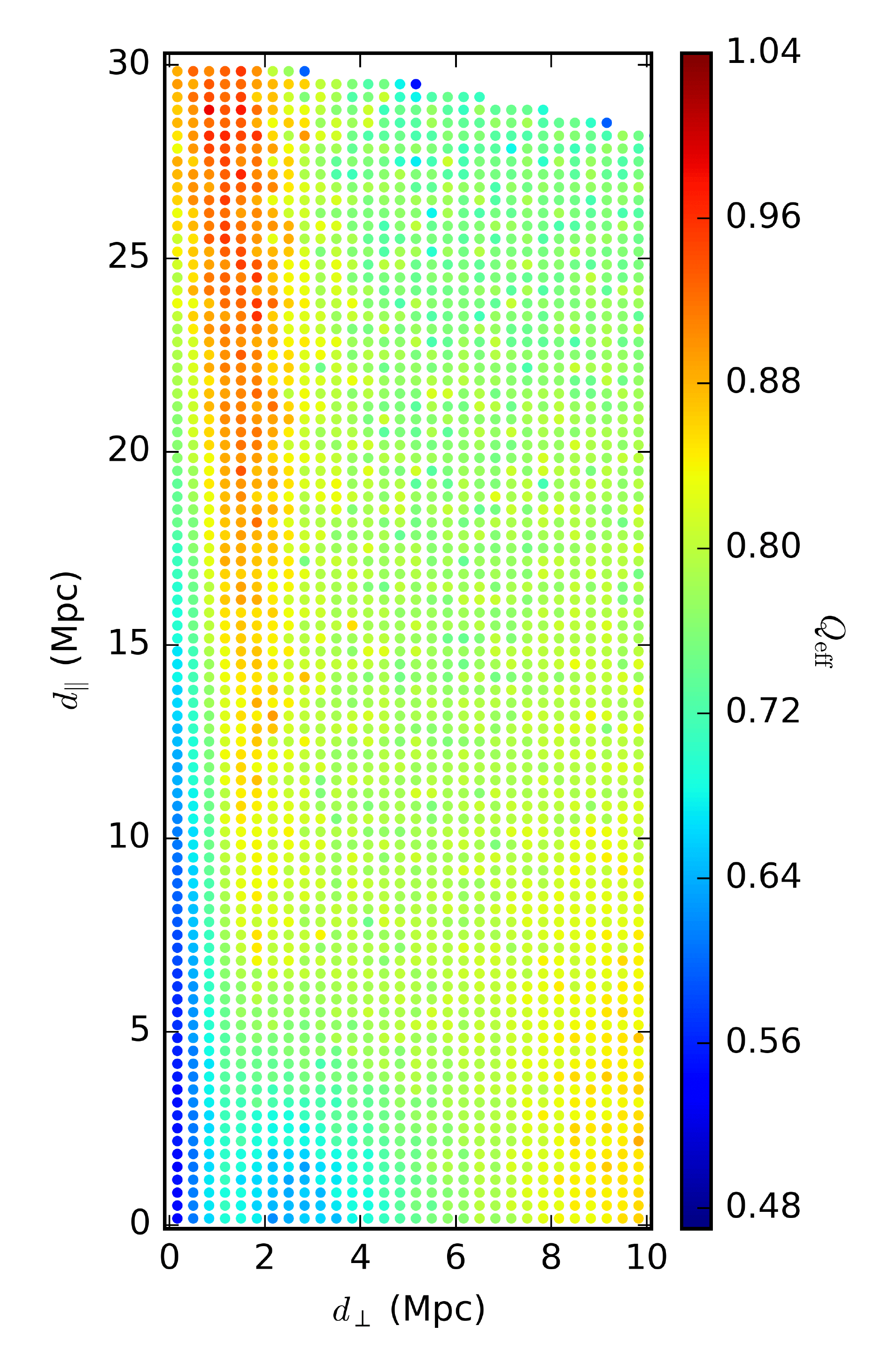}
\vspace{-0.6cm}
\caption{A zoom-in version of the $\qeff$ signal in redshift space. Specifically, we use finer bins on pair separation and zoom in on the $d_{\perp}$ axis so to be comparable to the real space $\qeff$ signal plotted in the bottom panel of Figure~\ref{fig:results_real}. To compensate for the loss in signal-to-noise due to using finer bins, we re-use each of the 16 different simulation boxes 16 times using different random seeds for the HOD.
The weighting function is the Gaussian-sigmoid function for $d_{p, \rm{max}} = 30$~Mpc as described in Section.~\ref{sec:weighting}.}
\label{fig:results_rsd_zoom}
\end{figure}

Similar to the last section, we can incorporate RSD and calculate the $\bpg$ and $\qeff$ in redshift space, as shown in Figure~\ref{fig:results_rsd}. The top panel shows the the pair-galaxy bias $\bpg$ as a function of pair separation. The middle panel shows the corresponding uncertainty on $\bpg$. The bottom panel shows the reduced 3PCF $\qeff$ in redshift space. For these plots, the weighting function is chosen to be the $d_{p, \rm{max}} = 30$~Mpc. Figure~\ref{fig:results_rsd_zoom} shows a zoom-in version of the $\qeff$ signal in redshift space with finer bins on pair separation to be comparable to the real space $\qeff$ signal plotted in the bottom panel of Figure~\ref{fig:results_real}. To compensate for the loss in signal-to-noise by choosing finer bins, we use 16 different seeds on top of 16 different simulation boxes, as discussed in Section~\ref{sec:catalogue}. One should disregard the high values along the outer circular edge of plots as these values are calculated from small sample of pairs and are thus dominated by noise. 

The top and bottom panel of Figure~\ref{fig:results_rsd} show a vertical ``ridge" at $d_{\perp} \approx 2$~Mpc in $\bpg$ and $Q_{\rm{eff}}$ that corresponds to the $\sim 2$~Mpc peak in the real space $\bpg$ and $\qeff$ signal (Figure~\ref{fig:results_real}). This ridge tracks the halo mass peak at $d_p \sim 2$~Mpc in Figure~\ref{fig:mhalo_dp}, but smeared out along the LOS by FoG. The strong clustering along the ridge all the way to $d_{\parallel} \sim 30$~Mpc suggests that the galaxy pair-wise peculiar velocities reach $\sim 2000$~km/s in the most massive halos. The zoom-in in Figure~\ref{fig:results_rsd_zoom} shows that the ridge is more clustered at larger LOS separation, which is due to the fact that larger LOS pair separation corresponds to larger peculiar velocities of the galaxies, which in turn corresponds to more massive halos. We also point out that the infall velocity of merging dark matter halos may also contribute to this ridge. However, we do not expect the infall velocity to frequently reach $\sim 2000$~km/s and speculate that the strongest $\qeff$ signal towards the top of the ridge should be completely dominated by FoG. 
 
We also see weaker anisotropy at large pair separation in redshift space. This is due to the RSD smearing out the filamentary structures, thus also randomizing the preferred direction given an LRG pair. The rise in bias value towards larger pair separation perpendicular to the LOS can be explained by the fact that more distant pairs along the same filament are better at picking out the preferred direction of the filament, which gives rise to higher bias given the anisotropy of the weighting function. 

Both the real space and redshift space $\bpg$ signal show a factor of 2 variation. This suggests that the hierarchical ansatz on the 3PCF does not hold in the squeezed limit. However, as we have discussed in this section, the variations in the squeezed 3PCF signal provides rich information on the galaxy-halo connection on the small scale and on the filamentary structure on the large scale. In the following section, we explore using the squeezed 3PCF as a diagnostics on the galaxy-halo connection. 

\section{HOD dependency of the 3PCF}
\label{sec:hod_depend}

We next discuss the relationship between the squeezed 3PCF and the HOD. Specifically, we explore the possibility of using the squeezed 3PCF to break degeneracies in the 2PCF and constrain the high mass end of the HOD. 

\subsection{Controlling the 2PCF with the HOD}
\label{sec:2pf_hod}

To investigate how the squeezed 3PCF breaks degeneracies in the 2PCF, we first need to understand how the HOD controls the 2PCF and devise HOD recipes that would change the 3PCF while maintaining the same 2PCF. First we attempt a linear emulator to model the 2PCF as a function of HOD parameters. As discussed in Section~\ref{sec:2PCF}, we study the projected 2PCF $w(d_{\perp})$ as given by equation~\ref{equ:2PCF_los_integrated_explicit}. To the first order, we can expand the projected 2PCF as
\begin{equation}
w_i(\textbf{p}) \approx w_i(\textbf{p}_0) + \sum_{j = 1}^4 w_{i, p_j}(p_j - p_{0,j}),
\label{equ:2pf_emulate_1st}
\end{equation}
where $\textbf{p} = \left[p_1, p_2, p_3, p_4\right] = \left[\log_{10} M_{\rm{cut}}, \log_{10} M_1, \sigma, \alpha \right]$ are the HOD parameters that we use to control the 2PCF. Note that we omitted parameter $\kappa$ because our tests show that $\kappa$ perturbs the 2PCF insignificantly for the variation amplitude that we are interested in. $\textbf{p}_0$ denotes the Z09 design of the HOD, as quoted in Section~\ref{sec:catalogue}. $w_i(\bf{p})$ denotes the value of the projected 2PCF integrated over the $i$-th bin in $d_{\perp}$. By choosing an uniform set of bins on the projected 2PCF, we form a vector $\left[w_i(\bf{p})\right]$ that holds the shape information of the projected 2PCF. $\left[w_{i, p_j}\right]$ records the matrix of the derivatives of the 2PCF integrated over the $i$-th bin, $w_i(\bf{p})$, against the $j$-th HOD parameter, $p_j$. We can determine each derivative $w_{i, p_j}$ by independently varying each parameter of the HOD and tracking the variation incurred onto $w_i(\bf{p})$. Once we have $\left[w_{i, p_j}\right]$, we solve for the set of HOD parameters $\bf{p}$ that minimizes $|\left[w_i(\bf{p})\right]-\left[w_i(\bf{p}_0)\right]|$. However, it turns out that the linear model does not produce a good emulator for the projected 2PCF. Our tests show that the linear approximation breaks down when varying $\alpha$ by more than $\sim 1\%$. 

Thus, we develop an emulator that incorporates second-order derivatives. 
To formulate a second order emulator of $\left[w_i(\bf{p})\right]$, we expand equation~\ref{equ:2pf_emulate_1st} to the 2nd order 
\begin{align}
w_i(\textbf{p}) \approx & w_i(\textbf{p}_0) + \sum_{j = 1}^4 w_{i, p_j}(p_j - p_{0,j}) \nonumber \\
& + \frac{1}{2}\sum_{j = 1}^4\sum_{k = j}^4 w_{i,p_j p_k}(p_j - p_{0,j})(p_k - p_{0,k}),
\label{equ:2pf_emulate_2nd}
\end{align}
where $w_i$ corresponds to the integrated value of the projected 2PCF over the $i$-th bin. $w_{i, p_j}$ denotes the first derivatives while $w_{i,p_j p_k}$ denotes the second derivatives. Since for each $i$, we have 4 $w_{i, p_j}$ and 10 $w_{i,p_j p_k}$ unknowns, we use a total of 14 different and independent evaluations of $\left[w_i(\bf{p})\right]$ to solve for all 14 derivatives.

Once we have solved for the first and second derivatives on the projected 2PCF, we have determined the full parametrized form of our $2^{\rm{nd}}$ order emulator for $\left[w_i(\bf{p})\right]$. Then we vary the HOD parameters by first varying $\alpha$ (we choose $\alpha$ because it controls the number of satellites) by $+5\%$ and $-5\%$ and then perturbing the remaining HOD parameters to find the $\left[w_i(\bf{p})\right]$ that best emulates $\left[w_i(\bf{p}_0)\right]$.
Specifically, we minimize $\chi^2$ with an error on $\delta w_i$ that scales with inverse sample volume. Considering that our full sample live in a sphere of diameter $d_{p, \rm{max}}$ in pair separation space, the error is given by $\sigma_i^2 = \sqrt{d_{p, \rm{max}}^2 - d_{\perp, i}^2}/d_{\perp, i}$ for an even binning in $d_{\perp,i}$.

For both $\delta \alpha = 0.05$ and $\delta \alpha = -0.05$, we find local minima of $\chi^2$ from an arbitrarily selected set of initial locations in the HOD parameter space. Then we visually select the set of HOD parameters that best emulates the Z09 2PCF. Table~\ref{tab:parameters} shows the resulting parameters of these best fit HODs. The minimization is done using the Powell algorithm \citep{1964Powell}, Nelder-Mead algorithm and the Broyden-Fletcher-Goldfarb-Shanno (BFGS) algorithm \citep{1970Shanno, 1970bShanno}, which yield consistent best fits. 
\begin{table}
\scalebox{1}{
\begin{tabular}{ c | c c c}
\hhline {====}
$\alpha$ & $\log_{10} M_{\rm{cut}}$ & $ \log_{10} M_1$ & $ \sigma$  \\ 
\hline
1+0.05 & 13.35--0.086 & 13.8--0.018 & 0.85--0.16 \\ 

1--0.05 & 13.35+0.070 & 13.8--0.024 & 0.85+0.093 \\ 
\hline 
\end{tabular} 
}
\caption{The HOD parameters corresponding to $\delta \alpha = \pm 0.05$ that best reproduce the Z09 design 2PCF.}
\label{tab:parameters}
\end{table}

\begin{figure}
\centering
 \hspace*{-0.8cm}
\includegraphics[width=3.8in]{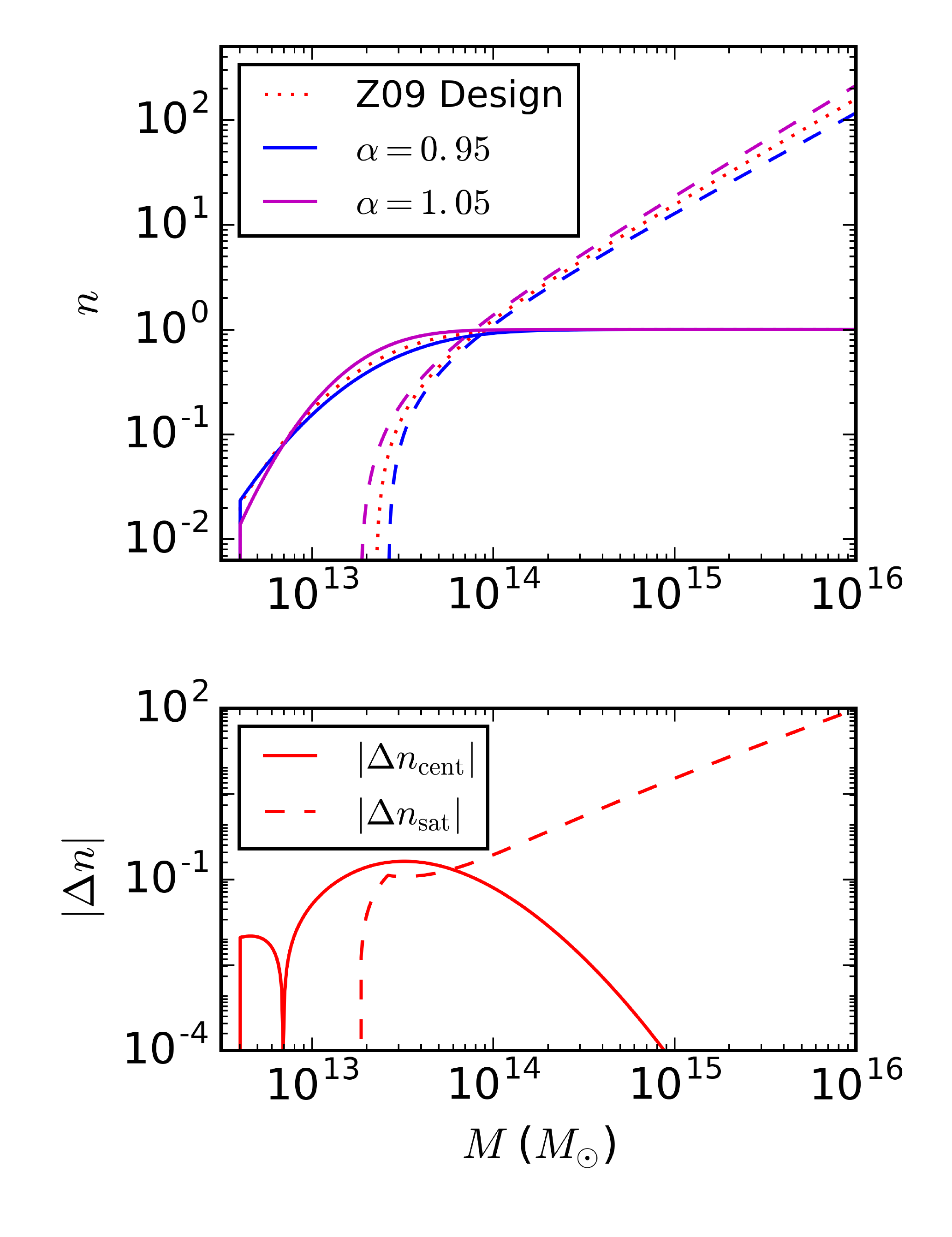}
\vspace{-1cm}
\caption{The HODs for $\alpha = 0.95$ (in blue) and $\alpha = 1.05$ (in magenta) that best fit the Z09 design HOD (in red). For the Z09 design, both $n_{\rm{cent}}$ and $n_{\rm{sat}}$ are plotted in dotted line. For the other two HODs, $n_{\rm{cent}}$ is plotted in solid curves whereas $n_{\rm{sat}}$ is plotted in dashed curves. The bottom panel shows the difference between the $\alpha = 1.05$ HOD and the $\alpha = 0.95$ HOD, where again the solid line shows the difference in $n_{\rm{cent}}$ and the dashed line shows the difference in $n_{\rm{sat}}$.}
\label{fig:3hod}
\end{figure}

We plot these two best fit HODs together with the Z09 design in Figure~\ref{fig:3hod}. The number of centrals and satellites are plotted in dotted red curve. The HOD corresponding to $\alpha = 0.95$ is plotted in blue and the HOD corresponding to $\alpha = 1.05$ is plotted in magenta. The solid curves show $n_{\rm{cent}}$ and the dashed curves show $n_{\rm{sat}}$. The bottom panel shows the absolute difference between the $\alpha = 1.05$ HOD and the $\alpha = 0.95$ HOD, where again the solid line shows the difference in $n_{\rm{cent}}$ and the dashed line shows the difference in $n_{\rm{sat}}$. We see that by increasing $\alpha$, we are increasing the number of satellites at the high-mass end of the HOD as expected. To compensate for such changes so as to maintain the same 2PCF, we see a decrease in the number of centrals at halo mass $M < 10^{13} M_{\odot}$. Thus, while we are increasing the number of LRG pairs in the largest halos, our algorithm is removing central-central pairs in the more common but smaller halos to compensate. 

\begin{figure}
\centering
 \hspace*{-0.6cm}
\includegraphics[width=3.6in]{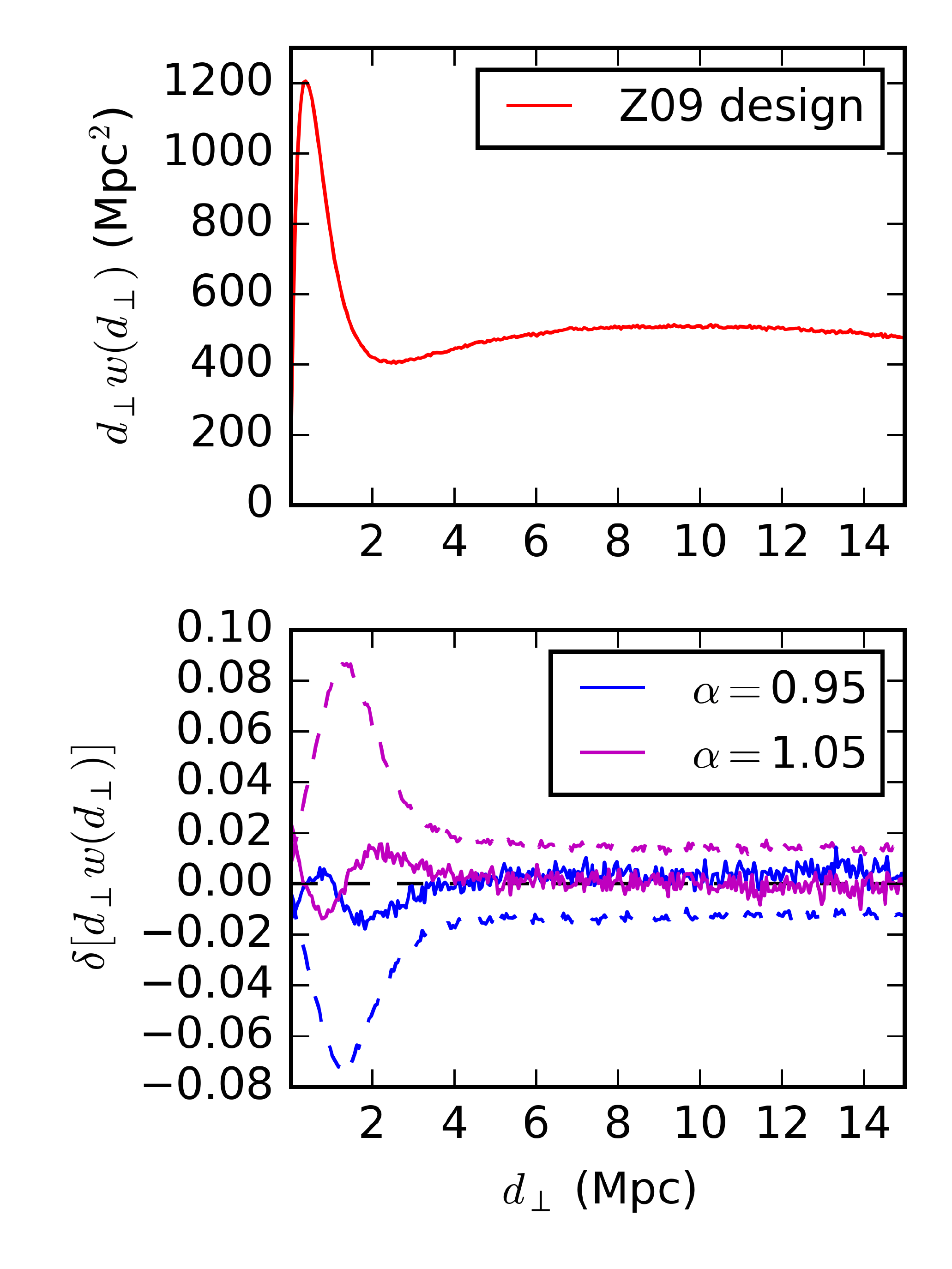}
\vspace{-0.8cm}
\caption{The top panel shows the projected 2PCFs corresponding to the Z09 design. The bottom panel show the change in projected 2PCFs corresponding to perturbed HODs, given in terms of $\Delta \left[d_{\perp} w(d_{\perp})\right] = d_{\perp} w_{\rm{perturbed}} - d_{\perp} w_{\rm{Z09}}$. The dashed curves correspond to only changing HOD parameter $\alpha$, with $\alpha = 0.95$ plotted in blue and $\alpha = 1.05$ plotted in magenta. 
The solid curves correspond to the best-fit HODs after perturbing the other HOD parameters to best reproduce the original 2PCF. The best fits suppress the difference between the $\alpha = 1.05$ case and the $\alpha = 0.95$ case from $< 16\%$ down to $< 3\%$. }
\label{fig:3-2pf}
\end{figure}

Figure~\ref{fig:3-2pf} top panel shows the projected 2PCF corresponding to the Z09 design. The bottom panel shows the relative change in the projected 2PCFs after perturbing HOD parameters. The dashed curves give the relative change in projected 2PCF after only changing $\alpha$, with blue corresponding to $\alpha = 0.95$ and magenta corresponding to $\alpha = 1.05$. The solid curves show the relative change in projected 2PCF after also perturbing the other HOD parameters to suppress the deviations in the projected 2PCF from that of the Z09 design. We see that for a $10\%$ change in $\alpha$, the best-fit 2PCFs shows an offset of $\sim 3\%$ at most, with the largest fractional deviations occurring between $d_{\perp} \approx 2$~Mpc. The fractional deviation is below $1\%$ beyond 4~Mpc. This change is considerably smaller than the $\sim 16\%$ change that occurs in the projected 2PCF before we adjust the other HOD parameters. 

\subsection{3PCF dependence on the HOD}
\label{sec:3pf_hod}

Now we compute the change in the reduced 3PCF $Q_{\rm{eff}}$ corresponding to the two best-fit HODs that we found in the last section. 

\begin{figure}
\centering
 \hspace*{-0.5cm}
\includegraphics[width=3.6in]{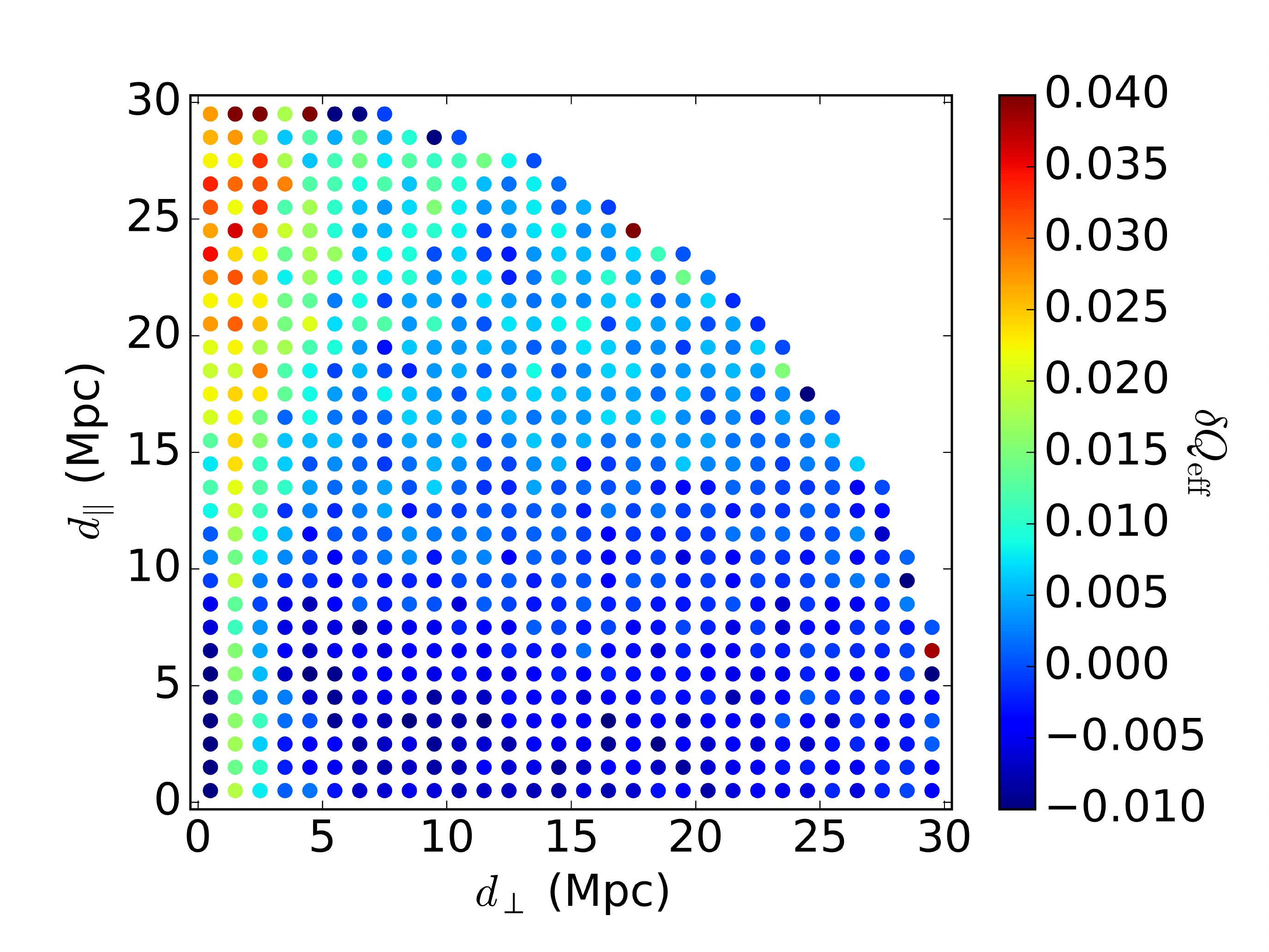}
\vspace{-0.5cm}
\caption{The relative change in reduced 3PCF $\delta Q_{\rm{eff}} = [Q_{\rm{eff}}(\alpha = 1.05) - Q_{\rm{eff}}(\alpha = 0.95)]/Q_{\rm{eff}}(\alpha = 1)$ plotted as a function of pair separation parallel to LOS $d_{\parallel}$ and pair separation perpendicular to LOS $d_{\perp}$. The $\qeff$ is given in redshift space.}
\label{fig:dQeff}
\end{figure}

\begin{figure}
\centering
 \hspace*{-0.5cm}
\includegraphics[width=3.6in]{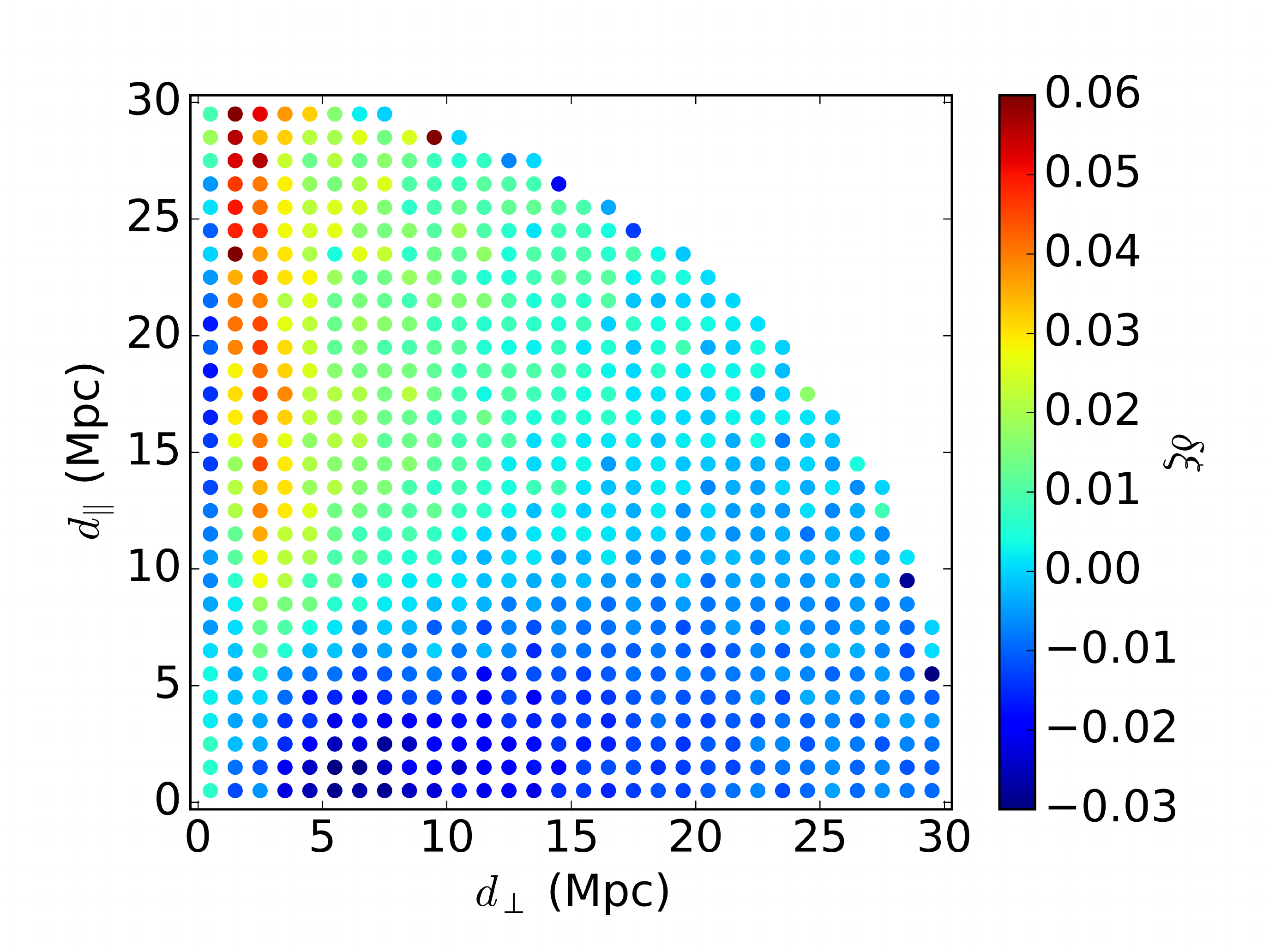}
\vspace{-0.5cm}
\caption{The change in the redshift-space 2PCF as a function of pair separation. Specifically, we define $\delta \xi(d_p) = (\xi(\alpha = 1.05)- \xi(\alpha = 0.95))/\xi(\alpha = 1)$.}
\label{fig:dxi}
\end{figure}

We plot the relative change in $Q_{\rm{eff}}$ in Figure~\ref{fig:dQeff}, where $\delta Q_{\rm{eff}}$ is defined as 
\begin{align}
\delta Q_{\rm{eff}} = \frac{\left[Q_{\rm{eff}}(\alpha = 1.05) - Q_{\rm{eff}}(\alpha = 0.95)\right]}{Q_{\rm{eff}}(\alpha = 1)}.
\end{align}
The $\delta \qeff$ signal is averaged over 16 simulation boxes, each with 16 different seeds to suppress shot noise.
We see a relative change in $Q_{\rm{eff}}$ not greater than $4\%$ at $d_{\perp}\sim$~1--3~Mpc, with the relative change increasing towards large $d_{\parallel}$. This suggests that $Q_{\rm{eff}}$ is most sensitive to $\delta \alpha$ along the 2 Mpc ridge (refer to the bottom panel of Figure~\ref{fig:results_rsd}) that corresponds to the most massive halos. 
This makes sense since $\alpha$ controls the number of galaxies in massive halos. 

In detail, the morphology of $\delta \qeff$ shown in Figure~\ref{fig:dQeff} recalls the variations in the projected 2PCF (Figure~\ref{fig:3-2pf}): an excess of $3\%$ at 2~Mpc scale, dominating at large $d_{\parallel}$ where the most massive halo pairs occur, followed by $<1\%$ differences at larger $d_{\perp}$.  We therefore wanted to check how the $\qeff$ variations in Figure~\ref{fig:dQeff} compared to the variations in the full anisotropic 2PCF. This is shown in Figure~\ref{fig:dxi}, where we have computed $\xi(d_{\perp},d_{\parallel})$ for both the $\alpha=0.95$ and $\alpha=1.05$ HODs and then shown the fractional difference. One sees that while the broadest patterns coincide, the details vary significantly. Specifically, the maximum variation in $\xi$ is $\sim 6\%$, twice as large as that of $\qeff$. We see a strong signal in $\delta \qeff$ at the smallest $d_{\perp}$ but not in $\delta\xi$. The ridge in $\delta \qeff$ extends all the way down to the smallest $d_{\parallel}$ whereas the $\delta \xi$ signal only extends down to about $d_{\parallel} \sim 6$~Mpc. There is also a consistently negative region in $\delta \xi$ at $d_{\perp}\sim$~5--15~Mpc, which is not reflected in the $\delta\qeff$ signal. The angular gradient at large pair-separation is also different between $\delta \qeff$ and $\delta \xi$. 
Thus, we conclude that the reduced 3PCF $\qeff(d_{\perp},d_{\parallel})$ is not a simple rescaling of the anisotropic 2PCF $\xi(d_{\perp},d_{\parallel})$. However, as the variations in the reduced 3PCF $\qeff$ ($<4\%$) are similar in size and scale to those in the projected 2PCF, it is unclear whether the squeezed 3PCF is breaking a degeneracy within this particular space of HOD models. 

While this test was inconclusive within this model space, we expect that there could be reasonable extensions of the HOD where the variation to $\qeff$ is disentangled from the variation to the projected 2PCF. For these scenarios, $\qeff$ would be a new diagnostic to distinguish among different HOD models. However, it is also possible that the squeezed 3PCF is well predicted by the 2PCF for certain classes of HOD models. Such a predicative relationship would be interesting in itself, as any discrepancy between the observed squeezed 3PCF and the 2PCF could then be used to falsify such classes of HOD models. 

In future works, we intend to apply our methodology to more sophisticated halo occupation models that incorporate dependence on more parameters than the one we used for this paper, such as models that incorporate halo assembly bias \citep{2002Wechsler, 
2005Gao, 2006Wechsler, 2007Croton, 2014Zentner, 2017Lehmann}. In particular, \citet{2016Hearin} parameterize the assembly bias in an HOD-like model (``Decorated HOD"). Recently, there has also been a series of studies exploring the dependence of the HOD on halo environment \citep[e.g.][]{2006Mandelbaum, 2011Gilmarin, 2012Croft, 
2015Tonnesen} and halo geometry \citep{2012vanDaalen}. It is plausible that the pair-galaxy bias and the squeezed 3PCF can offer extra constraints for these sophisticated halo occupation models. 

\section{Conclusions}
\label{sec:conclusions}
To summarize, this paper presents the methodology of cross-correlating the galaxy pair field with the galaxy field as a squeezed 3PCF. We develop a fast and efficient method using FFTs to compute the pair-galaxy bias $b_{p,g}$ and its associated reduced squeezed 3PCF $\qeff$.
We implement our method using a series of N-body cosmology simulations and show the $b_{p,g}$ and $\qeff$ as functions of pair separation in real space and in redshift space. In real space, a significant peak in $\bpg$ and $\qeff$ at pair separation $\sim$2~Mpc is observed, which is due to the fact that galaxy pairs at $\sim$2~Mpc separation trace the most massive dark matter halos. 
We also see strong anisotropy in the $b_{p,g}$ and $\qeff$ signals that tracks the large-scale filamentary structure. 

In redshift space, the FoG effect smears out the 2~Mpc peak into a ridge along the LOS. The anisotropy signal is also smeared out along the LOS, but persists at large pair separation. 
Both in real space and redshift, we see a factor of 2 variation in the squeezed 3PCF, contradicting the hierarchical ansatz. However, the variations in the squeezed 3PCF offers rich information on the galaxy-halo connection and the large-scale filamentary structure. 
 
Thus, we investigate using the reduced squeezed 3PCF $\qeff$ to constrain the high mass end of the HOD while breaking degeneracies in the 2PCF. We develop a second order emulator to model the projected 2PCF around the Z09 design HOD. Using the emulator, we find two simple HOD prescriptions that constrain the variations in the projected 2PCF to $<3\%$ when handed a $10\%$ perturbation to HOD parameter $\alpha$. The resulting perturbation to the squeezed 3PCF is no greater than $4\%$, which cannot be immediately disentangled from the variations in the 2PCF. Nevertheless, we propose that more sophisticated HOD models or different variations to the HOD could further disentangle the squeezed 3PCF from the 2PCF. 

In future papers, we intend to construct a more sophisticated emulator of the 2PCF as a function of HOD parameters. Another key objective is to extend our methodology to more sophisticated halo occupation models, such as ones that incorporate assembly bias, and explore how the pair-galaxy bias and the squeezed 3PCF can be used to distinguish among these more complex models. Eventually we want to apply our methodology to observations to detect these signals. 

\section*{Acknowledgements}
We would also like to thank Dr. Zachary Slepian and Joshua Speagle for useful discussions. DJE and LG are supported by National Science Foundation grant AST-1313285. DJE is additionally supported by U.S. Department of Energy grant DE-SC0013718 and as a Simons Foundation Investigator.
\bibliographystyle{mnras}
\bibliography{biblio}
\label{lastpage}
\end{document}